\documentclass[
  modern,      % "more-readable version" -- wider margins and bigger text
]{aastex63}  % [TODO] using aastex7 causes citations to break :(
    \usepackage{subfiles}
    \usepackage{comment}  % for multi-line comments, can use \comment .. \endcomment
    \usepackage{amsmath}
    \usepackage{graphicx}   % for images (e.g. in figures)
    \usepackage{gensymb}   % for degree symbol
    \usepackage{float}  % for figures HERE (use [H])
    \usepackage{xstring}  % for \IfSubStr
  \graphicspath{{./graphics/}}

    \newcommand{\customhalfwidth}{
      \ifdim 2\linewidth>\textwidth   % single-column mode
        0.5\textwidth
      \else   % two column mode
        \linewidth
      \fi
    }
    \newcommand{\uu}[1]{\ensuremath{\,\mathrm{#1}}}        % units, e.g. 5\uu{m/s}  -->  5 m/s
    \newcommand{\uue}[2]{\ensuremath{\,\mathrm{#1}^{#2}}}  % units with exponent, e.g. 5\uue{m}{2}  -->  5 m^2

    \newcommand{\advect}[2]{\frac{d_{#2} #1}{dt}}             % advective derivative of arg1, using fluid arg2.
    \newcommand{\pdtime}[1]{\frac{\partial #1}{\partial t}}   % partial derivative w.r.t. time.
    \newcommand{\inlineAdvect}[2]{d_{#2} #1 / dt}
    \newcommand{\inlinePdtime}[1]{\partial #1 / \partial t}
    \newcommand{\grad}[1]{\nabla #1}                          % gradient of arg1
    \newcommand{\divg}[1]{\nabla \cdot #1}                    % divergence of arg1
    \newcommand{\eqdef}{=}    % A \eqdef B  --> "A is defined to be B"  % this is the "boring" option, just '='.
    \newcommand{\enc}[3]{\left#1#2\right#3}
    
    \newcommand{\parens}[1]{\enc{(}{#1}{)}}
    \newcommand{\brackets}[1]{\enc{[}{#1}{]}}
    
    \newcommand{\textsup}[1]{^{\text{(#1)}}}  % superscript with (text)
    \newcommand{\textsub}[1]{_{\text{#1}}}    % subscript with text

    \newcommand{\W}{\omega}    %  ω     %useful in perturbation theory
    \newcommand{\K}{\vec{k}}   %  k     %useful in perturbation theory
    \newcommand{\oZ}{^{(0)}}  % goes after term to indicate zeroth order
    \newcommand{\Ldebye}[1]{\lambda_{D, #1}}

    \newcommand{\parensIfHasPlus}[1]{\IfSubStr{#1}{+}{(#1)}{#1}}  % if #1 has a '+' in it, put parentheses around it.
    \newcommand{\textIfInMathMode}[1]{\ifmmode\text{#1}\else#1\fi}  % if in math mode, use text mode for #1
    \newcommand{\iI}[1]{\textIfInMathMode{#1I}}  % ionI
    \newcommand{\iII}[1]{\textIfInMathMode{\parensIfHasPlus{#1}II}}  % ionII
    \newcommand{\subII}[1]{_\iII{#1}}  %ionI, as subscript
    \newcommand{\code}[1]{\texttt{\detokenize{#1}}}
    \newcommand{\BUCSP}{Boston University Center for Space Physics, 725 Commonwealth Ave, Boston, MA 02215, USA}

\begin{document}

%%% ---------------- TITLE / AUTHOR / ABSTRACT ---------------- %%%
\title{The Unstable Chromosphere: Effects of the Thermal Farley-Buneman Instability Across a Broad Range of Solar Chromospheric Conditions}

\author[0000-0002-1127-7350]{Samuel Evans}
  \affiliation{\BUCSP}
  %\email required for all authors if using aastex7. Unknown command in aastex63 though.
  %\email{sevans7@bu.edu}
\author[0000-0002-8581-6177]{Meers Oppenheim}
  \affiliation{\BUCSP}
\author[0000-0002-0333-5717]{Juan Mart\'inez-Sykora}
  \affiliation{SETI Institute, Mountain View, CA 94043, USA}
  \affiliation{Lockheed Martin Solar \& Astrophysics Laboratory, 3251 Hanover St, Palo Alto, CA 94304, USA}
  \affiliation{Rosseland Centre for Solar Physics, University of Oslo, P.O. Box 1029 Blindern, N-0315 Oslo, Norway}
  \affiliation{Institute of Theoretical Astrophysics, University of Oslo, P.O. Box 1029 Blindern, N-0315 Oslo, Norway}
\author[0000-0002-7951-9184]{Alexander Green}
  \affiliation{\BUCSP}

\begin{abstract}
  In the coldest regions of the solar atmosphere, lingering discrepancies between models and observations may be caused by the Thermal Farley–Buneman Instability (TFBI).
  This meter-scale, electrostatic, collisional, multifluid plasma instability converts energy from neutral flows into turbulent motions and heating.
  In the neutral frame of reference, these neutral flows manifest as an electric field which can drive the TFBI.
  In this work, we simulate the TFBI across a broad range of solar chromospheric conditions.
  We find clear proportionality between between TFBI-driven relative temperature increases of charged species ($\Delta T_{s}\textsup{turb} / T_{s}\oZ$) and driving electric field strength relative to the theoretical threshold field required for TFBI growth.
  We also discover a correlation between relative driving field strength and turbulent motions.
  Additionally, the TFBI consistently causes average current density to rotate towards the driving field, with Pederson/ambipolar component ($\vec{J} \cdot \hat{E}\oZ$) increasing by up to roughly 60\% while the Hall component ($\vec{J} \cdot \hat{E}\oZ \times \hat{B}$) decreases in magnitude by up to roughly 80\%.
  Meanwhile, we find TFBI-driven turbulence increases neutral heating rates by $(43\pm7)\%$ on average, with 85\% (28/33) of simulations having mean increase more than $29\%$, and a flat line of best fit suggesting zero correlation with driving field.
\end{abstract}

%%% ---------------- INTRODUCTION ---------------- %%%
\section{Introduction}
  \label{sec:intro}
  
  The chromosphere is the complex interface region between the solar surface and the million degree corona, a thin and irregularly shaped layer which plays an important role in energy transfer throughout the solar atmosphere.
    Accurately modeling this region remains challenging, with models struggling to predict temperatures and spectral line widths.
    Large-scale models predict temperatures too cold by more than 1000\uu{K} compared to the coldest temperatures associated with any observations of the Sun, often in simulated bubbles of plasma cooled by expansion 2000\uu{K} or less
      \citep[see, e.g.,][]{
      Wedemeyer2004,  % early large-scale simulation, 10^{3.25} K (~1780 K).
      Fontenla2009,  % 1d semi-empirical models
      Leenaarts2011,  % large-scale simulation, 1660 K
      Carlsson2016,  % large-scale simulation, artificial heating below 2500 K, reaches minimum allowed temperature of 2400 K.
      Loukitcheva2019,  % ALMA observations quoting minimum temperature of 4370 K.
      daSilvaSantos2020,  % inversions of ALMA & IRIS, 3000 K
      MartinezSykora2020.NEI,  % Bifrost simulation with NEI, less than 2000 K. (Also in Evans+2026.)
      Chintzoglou2021,  % ALMA & IRIS, and comparison to Bifrost.
      Mulay2021}.  % observe H2 at 4200 K from "temperature minimum region"
    Inferred temperatures from observations of comparable cold bubbles are closer to 3300\uu{K} or hotter
      \citep{Ortiz2014,  % cold bubble observations; Ca II / Fe I
      delaCruzRodriguez2015,  % cold bubble observations; Ca II
      Centeno2017,  % cold bubble observations, down to 3300 K.
      Stauffer2022,  % cold bubble observations; CO
      Song2023}.  % cold bubble observations (CO) & qualitatively similar MHD
    Synthesized profiles of \iII{Mg}~h~and~k lines also tend to be too narrow compared to those observed by the Interface Region Imaging Spectrograph (IRIS) \citep{DePontieu2014, Carlsson2019, Hansteen2023, Ondratscheck2024}.
    Difficulties synthesizing lines from models, incorporating all relevant physics into simulations, or inferring plasma parameters from observations all might contribute to these discrepancies between models and observations.
  
  Recent works predict Thermal Farley-Buneman Instability (TFBI) growth across a wide range of chromospheric parameters, show that the TFBI drives turbulent motions and heating, and discuss how large-scale models presently exclude TFBI effects \citep{Oppenheim2020, Dimant2023, Evans2023, Evans2025, Evans2026a}.
    Large-scale single-fluid MHD models exclude TFBI effects by default because they do not reach the relevant resolution, on the order of a few meters.
    Additionally, they neglect drifts between charged species \citep[see, e.g.,][]{Zaqarashvili2011, Khomenko2014, Shelyag2016, NobregaSiverio2020.AmbipolarBifrost}, an assumption which breaks down in weakly ionized regions of the chromosphere where electromagnetic forces govern electron motion while ions are dragged by collisions with flowing neutrals.
    The resulting drift between electrons and ions is precisely what drives the TFBI.
    Incorporating into models the combination of TFBI effects along with the zeroth-order physics of frictional heating associated with these neglected drifts \citep{Evans2026a} might be enough to yield large-scale models that closely match observations of the chromosphere.
  
  As noted in \citet{Evans2026a}, quantifying the effects of the TFBI across the chromosphere requires two essential components:
    (1) a map of where the TFBI occurs across the chromosphere, and (2) some way to predict the amount of TFBI-driven turbulent heating and turbulent motions as a function of physical parameters across the chromosphere.
    While that work addressed the first point, this work instead focuses entirely on point (2).
    Prior studies showed that the TFBI can occur within chromospheric parameters \citep{Oppenheim2020, Evans2023}, and that its turbulent effects are reasonably robust to modeling methodology and numerical parameters across multifluid 2D, kinetic 2D, and kinetic 3D simulations of the same physical parameter regime \citep{Evans2025}.
    Here, we perform a suite of TFBI simulations across a wide range of chromospheric parameters, quantify the resulting turbulent motions and heating, and discover trends in the results.
    This ultimately enables us to predict TFBI effects across the chromospheric parameter regime.
  
  The remainder of this paper is structured as follows.
    Section~\ref{sec:methods} discusses relevant physical equations and assumptions, describes the EPPIC simulator used throughout this work, and presents the TFBI suite simulation parameters.
    Section~\ref{sec:results:typical_tfbi} analyzes a typical example from the TFBI simulation suite, while
    Section~\ref{sec:results:tfbi_effects_across_chromo} quantifies the effects of TFBI turbulence across the chromospheric parameter regime.
    Finally, Section~\ref{sec:discussion} discusses the potential applications and limitations of these results, while Section~\ref{sec:conclusion} concludes by highlighting the main results.

%%% ---------------- METHODS ---------------- %%%
\section{Methods}
  \label{sec:methods}
  
  This work simulates the TFBI across a broad range of conditions throughout the Sun's chromosphere.
  Section~\ref{sec:methods:governing_equations} describes the multifluid equations relevant to the TFBI.
  Section~\ref{sec:methods:eppic_simulator} describes the kinetic simulator, EPPIC, used throughout this work to simulate the TFBI.
  Section~\ref{sec:methods:tfbi_suite_params} details the broad range of parameters used throughout the suite of TFBI simulations in this work.
  
  \subsection{Governing Multifluid Equations}
    \label{sec:methods:governing_equations}
  
    Although multifluid equations exclude kinetic effects, they still lead to useful predictions about the TFBI, such as its predicted growth rate and wavelength \citep[see e.g.,][]{Evans2026a}.
    Prior work has also demonstrated the possibility for close agreement between kinetic and multifluid TFBI simulations \citep{Evans2025}.
  
    Excluding kinetic effects, one may consider a multifluid plasma with number densities ($n_{s}$), velocities ($\vec{u}_{s}$) and temperatures ($T_{s}$) governed by the continuity, momentum, and thermal equations for each charged species, $s$:
    \begin{subequations} \label{eqs:governing}
      %% EXTRA SPACING %%
        % \begin{equation} \label{eq:continuity}
        %   \pdtime{n_{s}} + \divg{\parens{n_{s} \vec{u}_{s}}} = 0
        % \end{equation}
        % \begin{equation} \label{eq:momentum}
        %   \advect{\vec{u}_{s}}{s}
        %   =
        %   - \frac{\grad{\parens{n_{s} k_B T_{s}}}}{m_{s} n_{s}}   % replaced P = n kB T.
        %   + \frac{q_{s}}{m_{s}} \parens{ \vec{E} + \vec{u}_{s} \times \vec{B} }
        %   - \nu_{s,n} \vec{u}_{s}
        % \end{equation}
        % \begin{equation} \label{eq:heating}
        %   \advect{T_{s}}{s}
        %   =
        %   - \frac{2}{3} T_{s} \divg{\vec{u}_{s}}
        %   + \frac{2 m_{s}}{m_{s} + m_{n}} \nu_{s,n} \brackets{
        %       \frac{m_{n}}{3 k_B} |\vec{u}_{s}|^2 + \parens{T_{n} - T_{s}}
        %     }
        % \end{equation}
      %% APJ SPACING (remove all unnecessary internal spaces) %%
      \begin{equation} \label{eq:continuity}
        \pdtime{n_{s}}+\divg{\parens{n_{s}\vec{u}_{s}}}=0
      \end{equation}
      \begin{equation} \label{eq:momentum}
        \advect{\vec{u}_{s}}{s}=-\frac{\grad{\parens{n_{s}k_{B}T_{s}}}}{m_{s}n_{s}}+\frac{q_{s}}{m_{s}}\parens{\vec{E}+\vec{u}_{s}\times\vec{B}}-\nu_{s,n}\vec{u}_{s}
      \end{equation}
      \begin{equation} \label{eq:heating}
        \advect{T_{s}}{s}=-\frac{2}{3}T_{s}\divg{\vec{u}_{s}}+\frac{2m_{s}}{m_{s}+m_{n}}\nu_{s,n}\brackets{\frac{m_{n}}{3k_{B}}|\vec{u}_{s}|^{2}+\parens{T_{n}-T_{s}}}
      \end{equation}
      where $\inlineAdvect{f}{s}\eqdef\inlinePdtime{f}+\vec{u}_{s}\cdot\grad{f}$; the Boltzmann constant is $k_{B}$; and all variables, including the electric field ($\vec{E}$), are expressed in the neutral reference frame ($\vec{u}_{n}=0$).
      The variables $m_{s}$ and $q_{s}$ represent the mass and signed charge of species $s$ (with $q_{s}<0$ for electrons), while $m_{n}$ and $T_{n}$ represent the mass and temperature of neutrals.
      This work considers only neutral hydrogen, due to its prevalence throughout the Sun's chromosphere.
      The collision frequencies $\nu_{s,n}$ model charged-neutral collisions, while the equations neglect charged-charged collisions (Coulomb collisions) which are usually orders of magnitude smaller in the lower-to-mid chromosphere below $10,000$~K \citep[][and references therein]{Wargnier2022}.
  
      Meanwhile, the system is assumed to be electrostatic, with constant magnetic field ($\inlinePdtime{\vec{B}}=0$), while the electric field is governed by Gauss's law:
      \begin{equation} \label{eq:gauss}
        % \divg{\vec{E}} = \frac{1}{\epsilon_0} \sum_{s} q_{s} n_{s}  % extra spacing
        \divg{\vec{E}}=\frac{1}{\epsilon_{0}}\sum_{s}q_{s}n_{s}  % apj spacing
      \end{equation}
      where $\epsilon_0$ is the vacuum permittivity constant.
    \end{subequations}
  
    While linear TFBI theory assumes weakly ionized plasma, with effectively constant neutral density, velocity, and temperature, in reality the neutrals will heat over time due to collisions with charged species.
    Note that the TFBI assumption of constant temperature is still reasonable as long as the heating timescale is much longer than the TFBI growth rate.
    To compute the neutral heating rate, the neutral thermal equation can be expressed in the neutral reference frame, analogously to equation~\eqref{eq:heating} but here summing contributions from collisions with all charged species and utilizing $\vec{u}_{n}=0$ to simplify derivatives:
    \begin{equation} \label{eq:dTndt}
      %% EXTRA SPACING %%
        % \pdtime{T_{n}} = \sum_s \frac{2 m_{s}}{m_{n} + m_{s}} \frac{n_{s}}{n_{n}} \nu_{s,n} \brackets{
        %       \frac{m_{s}}{3 k_B} |\vec{u}_{s}|^2 + \left( T_{s} - T_{n} \right)
        %       }
      %% APJ SPACING %%
      \pdtime{T_{n}}=\sum_{s}\frac{2m_{s}}{m_{n}+m_{s}}\frac{n_{s}}{n_{n}}\nu_{s,n}\brackets{\frac{m_{s}}{3k_{B}}|\vec{u}_{s}|^{2}+\left(T_{s}-T_{n}\right)}
    \end{equation}
      where we have used $m_{s}n_{s}\nu_{s,n}=m_{n}n_{n}\nu_{n,s}$, which comes from momentum conservation.
  
    Dropping derivatives and rearranging equations~\eqref{eq:momentum} and \eqref{eq:heating} yields zeroth-order solutions for $\vec{u}_{s}$ and $T_{s}$ \citep[see e.g.,][]{Dimant2023}:
    \begin{subequations} \label{eqs:zeroth_order}
      %% EXTRA SPACING %%
        % \begin{equation} \label{eq:u0}
        %   \vec{u}_{s}\oZ =
        %   \frac{1}{1 + \kappa_{s}^2}
        %   \brackets{
        %     \frac{\kappa_{s}}{B} \vec{E}\oZ   % pederson
        %     + \frac{\kappa_{s}^2}{B^2} \vec{E}\oZ\times\vec{B}  % hall
        %   }
        % \end{equation}
        % \begin{equation} \label{eq:T0}
        %   T_{s}^{(0)} = T_{n}
        %               + \frac{m_{n}}{3 k_B}
        %                 \frac{\kappa_{s}^2}{1 + \kappa_{s}^2}
        %                 \frac{|\vec{E}\oZ|^2}{B^2}
        % \end{equation}
      %% APJ SPACING %%
      \begin{equation} \label{eq:u0}
        \vec{u}_{s}\oZ=\frac{1}{1+\kappa_{s}^{2}}\brackets{\frac{\kappa_{s}}{B}\vec{E}\oZ+\frac{\kappa_{s}^{2}}{B^{2}}\vec{E}\oZ\times\vec{B}}
      \end{equation}
      \begin{equation} \label{eq:T0}
        T_{s}^{(0)}=T_{n}+\frac{m_{n}}{3k_{B}}\frac{\kappa_{s}^{2}}{1+\kappa_{s}^{2}}\frac{|\vec{E}\oZ|^{2}}{B^{2}}
      \end{equation}
    \end{subequations}
    where we have imposed the additional assumption $\vec{E}\oZ\cdot\vec{B}=0$, defined $B=|\vec{B}|$, and introduced the magnetization parameter:
    \begin{equation} \label{eq:kappas}
      % \kappa_{s} \eqdef \frac{q_{s} B}{m_{s} \nu_{s,n}}  % extra spacing
      \kappa_{s}\eqdef\frac{q_{s}B}{m_{s}\nu_{s,n}}  % apj spacing
    \end{equation}
    These zeroth-order solutions predict, for charged species, the background velocities and temperatures to which the system would settle over long enough timescales, in the absence of a plasma instability or other higher-order effects.
    Such timescales are easily less than 1~ms throughout the chromosphere, far faster than chromospheric dynamical timescales (a few seconds or more).  % if reviewer asks for proof: we can add, "they are less than 1 ms when evaluated across the TFBI-appropriate region of the chromosphere from Evans+2026.
    Thus, we set species' initial velocities and temperatures to $\vec{u}_{s}\oZ$ and $T_{s}\oZ$ at the start of each simulation run throughout this work.
  
    Combining the zeroth-order solutions for $\vec{u}_{s}\oZ$ and $T_{s}\oZ$ with the neutral thermal equation~\eqref{eq:dTndt} yields the ``zeroth-order'' neutral heating rate:
    \begin{equation} \label{eq:dTndt0}
      %% EXTRA SPACING %%
        % \parens{\pdtime{T_{n}}}\oZ = \sum_s \frac{2 m_{s}}{3 k_B} \frac{n_{s}\oZ}{n_{n}} \nu_{s,n} 
        %                       \frac{\kappa_{s}^2}{1 + \kappa_{s}^2}
        %                       \frac{|\vec{E}\oZ|^2}{B^2}
      %% APJ SPACING %%
      \parens{\pdtime{T_{n}}}\oZ=\sum_{s}\frac{2m_{s}}{3k_{B}}\frac{n_{s}\oZ}{n_{n}}\nu_{s,n}\frac{\kappa_{s}^{2}}{1+\kappa_{s}^{2}}\frac{|\vec{E}\oZ|^{2}}{B^{2}}
    \end{equation}
    Existing large-scale models of the solar atmosphere likely fail to properly implement this zeroth-order neutral heating rate, which may have a large impact across weakly ionized regions throughout the chromosphere \citep{Evans2026a}.
    %Furthermore, as shown later in this work (Section~\ref{sec:results:tfbi_effects_across_chromo}), the TFBI also causes heating, and we suggest large-scale models incorporate its effects by implementing a modified version of equation~\eqref{eq:dTndt0}.

  \subsection{The EPPIC Simulator}
    \label{sec:methods:eppic_simulator}
  
    In this work, we simulate the TFBI across a broad range of chromospheric parameters using the kinetic code, EPPIC.
    This section provides an overview of the EPPIC simulator and some of our choices of inputs directly related to the underlying numerical methods.
    Meanwhile, Section~\ref{sec:methods:tfbi_suite_params} and Table~\ref{tab:tfbi_suite_params} assign unique labels to each simulation and describe our choices of physical parameters as well.
    The EPPIC code itself is further described in \citet{Oppenheim2004}.
  
    EPPIC is a kinetic particle-in-cell simulator, which models each charged species as a collection of particles, while storing information about electric field in a grid of cells across space.
    The grid can be 3D, however previous work simulating the TFBI in a single parameter regime across different models found only slightly more heating in 3D than in 2D \citep{Evans2025}, so we restrict to only 2D simulations in this work.
    Each particle has a mass, charge, position, and velocity.
    Positions can take any value (on a continuum, to within floating point double precision) within the bounds of the 2D grid.
    Velocities have $x$, $y$, and $z$ components, though the $z$ component tends to remain relatively small for 2D simulations, which have no $z$ axis.
    
    Each timestep, positions change according to particle velocities; velocities change due to collisions with neutrals and the Lorentz force; and the electric potential ($\phi$) is updated based on Gauss's law~\eqref{eq:gauss}, using $\vec{E}=-\grad{\phi}$, after computing densities within each grid cell.
    Densities within each grid cell are computed by adding contributions from nearby particles, weighted appropriately considering each particle's distance from the grid cell center, via a linear tent function spanning one grid cell width \citep{Birdsall1991}.
    Each EPPIC simulation in this work uses a constant collision frequency for each species, treating the neutrals as a uniform stationary background of neutral hydrogen with $\vec{u}_{n}=0$.
    These simulations also add a constant, externally imposed electric field when calculating particle accelerations, to model the background electric field formed by chromospheric dynamics occurring at spatial and temporal scales much larger than the simulation itself.
  
    Some EPPIC inputs are directly related to the underlying numerical methods.
    The underlying noise levels in EPPIC tend to be inversely proportional to the square root of the number of PIC particles.
    To balance computational cost with reducing artificial particle noise, we choose to use 100 PIC electrons per cell in almost all cases.
    EPPIC also enables applying a low-pass filter to the electric potential to reduce grid scale fluctuations, with $k$-space filter width controlled by the $f\textsub{width}$ parameter; we choose $f\textsub{width}=3$ in almost all cases.
    For two simulations (labeled (2,1,3,3,6) and (3,0,3,2,4) in Table~\ref{tab:tfbi_suite_params}), we instead use 20 PIC electrons per cell to reduce computational cost, and $f\textsub{width}=9$ to cause initial temperatures more closely comparable to the equilibrium temperatures predicted by equation~\eqref{eq:T0}.
    \citet{Evans2025} (Appendix~A) further discusses the effects of varying the number of PIC particles and filter width.
    In all cases, for each ion species, we use between 1.6 to 12.5 times more PIC ions than PIC electrons.
    In particular, we use more PIC ions for species which permit more aggressive subcycling due to having larger timescales than electrons.
    The exact number of PIC ions and subcycling parameters chosen for each simulation can be found via Appendix~\ref{sec:appendix:repro}.
  
    Choosing appropriate cell widths and timesteps for simulations across a wide range of physical parameters is a nontrivial task.
    Setting grid cell widths too much larger than the Debye length can lead to numerical heating in kinetic PIC codes.
    For all simulations in this work, we choose grid cell width 10\% larger than the system's initial Debye length:
    \begin{equation} \label{eq:eppic_dx}
      %% EXTRA SPACING %%
        % \Delta x = 1.1 \ \lambda_{D,\text{tot}}\oZ
        %   ,\quad \text{where} \quad
        % \lambda_{D,\text{tot}}\oZ = \parens{\sum_{s} (\Ldebye{s}\oZ)^{-2}}^{-1/2}
      %% APJ SPACING %%
      \Delta{x}=1.1\ \lambda_{D,\text{tot}}\oZ,\quad\text{where}\quad\lambda_{D,\text{tot}}\oZ=\parens{\sum_{s}(\Ldebye{s}\oZ)^{-2}}^{-1/2}
    \end{equation}
    with $\Ldebye{s}\oZ$ defined in the usual way: $(\Ldebye{s}\oZ)^2=\epsilon_{0}k_{B}T_{s}\oZ/(n_{s}\oZ{q}_{s}^2)$.
    Meanwhile, if the timestep is larger than any relevant physical timescale, EPPIC might fail to properly model the corresponding physics.
    For example, any particles moving more than one or two cells per timestep will miss the chance to interact with fields in cells between their initial and final positions.
    For almost all simulations in this work, we choose timestep 10\% smaller than the smallest relevant initial timescale:
    \begin{equation} \label{eq:eppic_dt}
      %% EXTRA SPACING %%
        % \Delta t = 0.9 \ \min \brackets{
        %   2 \pi \omega_{p,s}^{-1},\
        %   2 \pi \Omega_{s}^{-1},\ 
        %   \nu_{s,n}^{-1},\ 
        %   \Delta x / v_{\text{therm},s},\ 
        %   \Delta x / (|\vec{E}\oZ|/B)
        % }
      %% APJ SPACING %%
      \Delta{t}=0.9\ \min\brackets{2\pi\omega_{p,s}^{-1},\ 2\pi\Omega_{s}^{-1},\ \nu_{s,n}^{-1},\ \Delta{x}/v_{\text{therm},s},\ \Delta{x}/(|\vec{E}\oZ|/B)}
    \end{equation}
    where the minimum is taken across all variables and charged species, though electrons happen to have the minimum timescale for each fluid-dependent variable across all simulations in this work.
    The variables correspond, respectively, to the timescales of:
      plasma oscillations, where the plasma frequency is $\omega_{p,s}=\sqrt{n_{s}\oZ{q}_{s}^2/(m\epsilon_0)}$;
      gyration, where the gyrofrequency is $\Omega_{s}=|q_{s}B/m_{s}|$;
      collisions with neutrals;
      thermal motion, where the thermal velocity is $v_{\text{therm},s}=\sqrt{k_{B}T_{s}\oZ/m_{s}}$;
      and the $|\vec{E}\oZ|/B$~speed, which serves as an upper bound for the zeroth-order drift speeds $|\vec{u}_{s}\oZ|$, as per equation~\eqref{eq:u0}.
    The smallest timescale is usually set by the electron thermal motion, however in two cases (labeled (3,1,4,2,6) and (3,3,3,2,4) in Table~\ref{tab:tfbi_suite_params}) it is instead set by the electron-neutral collision frequency, and in one case ((3,0,3,4,6)C) it is set by the electron gyrofrequency.
    Only one simulation ((3,1,3,3,6)) does not follow equation~\eqref{eq:eppic_dt} exactly, instead reducing timestep by an extra factor of 3, to avoid particles moving too far per timestep after turbulence significantly increases temperatures in some regions.

  \subsection{TFBI Simulation Suite Parameters}
    \label{sec:methods:tfbi_suite_params}
  
    This section describes the TFBI simulation suite: a group of 33 EPPIC simulations of the TFBI across a broad range of chromospheric parameters.
    In particular, this section details the input parameters for each simulation, as tabulated in Table~\ref{tab:tfbi_suite_params}.
  
    \begin{table}[htbp]
      \centering
      \includegraphics[width=\textwidth]{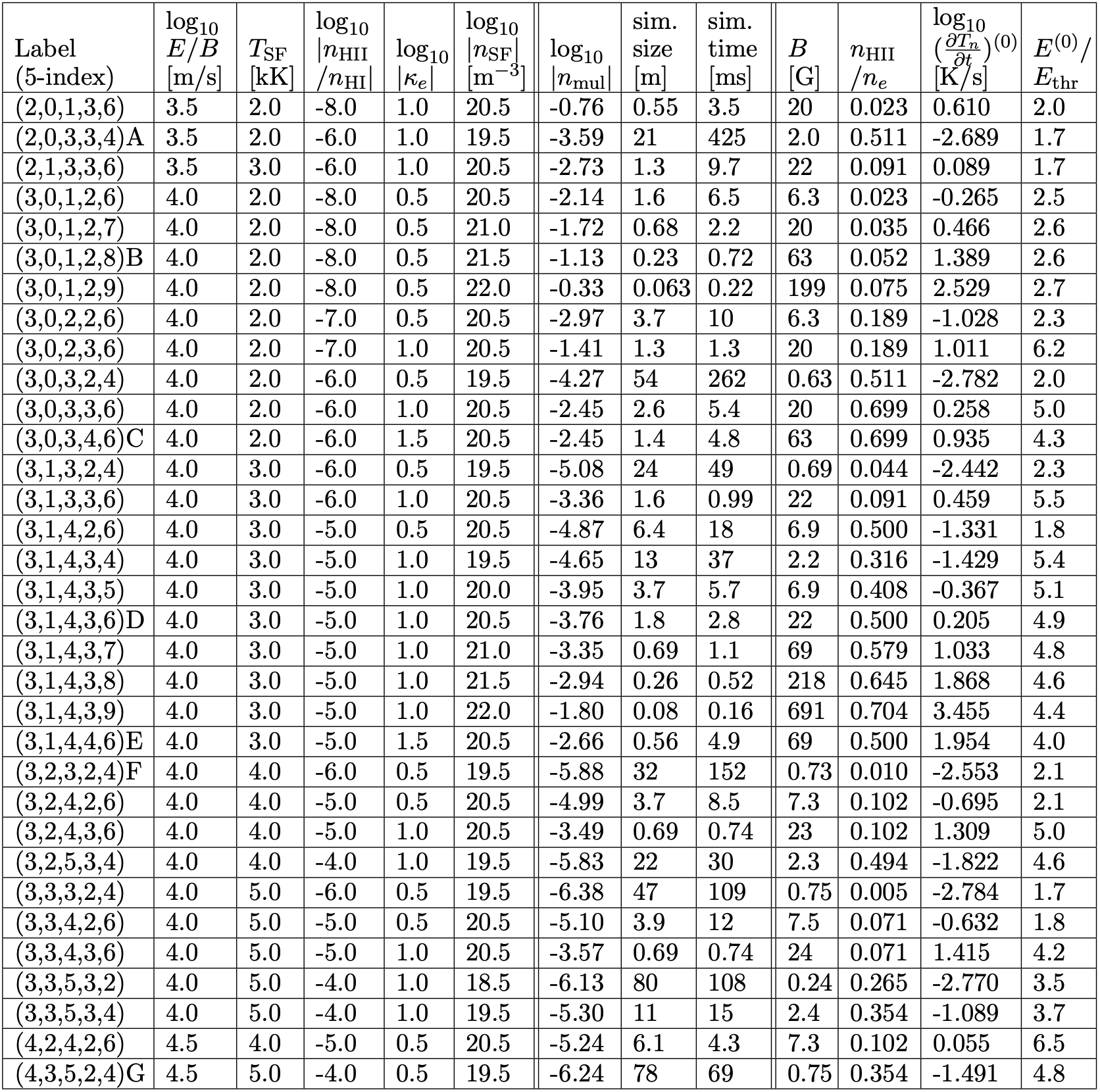}
      \caption{
        Parameters for each simulation in the TFBI simulation suite.
        The 5-index (first column, in parentheses) indicates the next five columns' values' indexes within the 5D grid from \citet{Evans2026a}.
        Labels end with a letter for all simulations mentioned explicitly later in the text, to facilitate finding them quickly in the table.
        The seventh column indicates the artificial number density scaling factor (via $\log_{10} |n\textsub{mul}|$) by which electron and ion densities were multiplied.
        The next two columns indicate simulation extent in space (side length of the 2D square simulation box) and time (simulation duration).
        The remaining columns provide values derived from the other input parameters:
          magnetic field strength ($B$),
          relative \iII{H} density ($n\subII{H}/n_{e}$),
          zeroth-order neutral heating rate (via $\log_{10}|(\inlinePdtime{T_{n}})\oZ~[K/s]|$) computed from equation~\eqref{eq:dTndt0},
          and ratio between electric field and threshold electric field required for TFBI growth (via $E\oZ/E\textsub{thr}$).
      }
      \label{tab:tfbi_suite_params}
    \end{table}
  
    Each simulation chooses a different combination of values for five independent variables, as tabulated by Table~\ref{tab:tfbi_suite_params}:
      $E/B$~speed,
      single-fluid temperature ($T_{SF}$),
      hydrogen ionization fraction ($n\subII{H}/n\textsub{H}$),
      electron magnetization ($\kappa_{e}$),
      and single-fluid number density ($n_{SF}$).
    The selected values of each variable occur at points in the 5D grid detailed in \citet{Evans2026a}, and the first column of Table~\ref{tab:tfbi_suite_params} shows each simulation's 5-index, which indicates each variable's value's index in the 5D grid.
    All available values are:
      $E/B=10^{2.5},10^{3.0},...,10^{5.0}\uu{m/s}$; 
      $T_{SF}=2000,3000,4000,5000\uu{K}$; 
      $n\subII{H}/n\textsub{H}=10^{-9.0},10^{-8.0},...,10^{-3.0}$; 
      $\kappa_{e}=10^{-0.5},10^{0.0},...,10^{2.5}$; 
      and $n_{SF}=10^{17.5},10^{18.0},...,10^{23.5}\uue{m}{-3}$.
  
    Multi-species inputs to EPPIC are derived from these five independent variables as outlined here and further detailed in \citet{Evans2026a}.
    Compute ion and neutral densities by
      plugging $T_{SF}$ and $n_{SF}$ into lookup tables to find $n_{e}$;
      assuming photospheric abundances \citep{Gustafsson1975} to compute densities of elements;
      plugging $n_{e}$ and $T_{SF}$ into the Saha equation to compute ionization fractions for non-hydrogen elements, neglecting any multiply-ionized species;
      combining the density and ionization fraction for each element to compute ion and neutral densities;
      and then neglecting all neutrals except hydrogen neutrals.
    Collision frequencies ($\nu_{s,n}$) for $s=e$ and $\iII{H}$ depend on masses, $T_{SF}$, $n_{n}$, and cross-sections from tables dependent on $T_{SF}$.
    For all other species, assume Maxwell molecule collisions, providing collision frequencies depending only on masses and $n_{n}$.
    Compute the magnetic field ($B$) from $\kappa_{e}$ and $\nu_{e,n}$ by rearranging equation~\eqref{eq:kappas}, then multiply by the input $E/B$~speed to compute $E$ in the neutral frame, $\vec{u}_n=0$.
    Set neutral temperature equal to the single-fluid temperature, $T_{n}=T_{SF}$, which should be reasonable for the weakly ionized plasma considered here.
    Set all charged species' initial velocities and temperatures to their zeroth-order equilibrium values from equations~\eqref{eqs:zeroth_order}.
    Finally, compute the initial electron number density from the charge-weighted sum of all ion number densities: $n_{e}=\sum_{i}q_{i}n_{i}/|q_{e}|$.
  
    Including more than five or six ions proved computationally challenging for the TFBI linear theory solver code, \code{tfbi_theory} \citep{Evans2026a, tfbi_theory_and_SymSolver}.
    Instead, we combine heavy ions with similar mass into groups, and treat each heavy ion group as a single species when computing TFBI theoretical predictions and when simulating it in EPPIC.
    The ion groups are: \iII{C+N+O+Ne}, \iII{Na+Mg+Al+Si+S}, and \iII{K+Ca+Cr+Fe+Ni}.
    The number density of each heavy ion group equals the sum of number densities of ions in the group.
    The mass of each heavy ion group equals the number-density-weighted average mass of each ion in the group.
    Additionally, \iII{He} is neglected entirely, as its contributions are negligible throughout the weakly ionized regions of the chromosphere.
    These groupings and choices about ions agree with those of \citet{Evans2026a}.
  
    Simulating the TFBI when directly using chromospheric number densities becomes extremely computationally expensive for EPPIC, which must resolve length scales close to the Debye length.
    However, scaling all charged species' number densities by the same constant factor has very little impact on the governing equations~\eqref{eqs:governing}.
    The continuity, momentum, and thermal equations are entirely unaffected by such scaling.
    Gauss's law is affected, however the impact on TFBI physics remains minor as long as ion collisional mean free paths remain larger than ion Debye lengths \citep{Rosenberg1998, Dimant2004, Dimant2023, Evans2026a}.
    Shrinking charged species' densities increases Debye lengths, but shrinking densities too much suppresses the TFBI, reducing growth rates and increasing the dominant wavelength where the maximum growth rate occurs \citep[see, e.g., Appendix~C of][]{Evans2026a}.
  
    The seventh column of Table~\ref{tab:tfbi_suite_params} indicates the charged species density scaling factor ($n\textsub{mul}$) applied to each simulation.
    The factors were chosen by hand, with the goal of minimizing computational cost while also not altering TFBI physics too much.
    Frequently, the chosen scaling factor simply minimizes the computational cost, as this occurs when increases in cost associated with suppressing the TFBI (causing larger required spatial and temporal simulation extent) become significant enough to outweigh decreases in cost associated with larger Debye length (causing larger permitted $\Delta{x}$).
    The full set of diagnostics used when choosing the density scaling factor is available for each simulation via Appendix~\ref{sec:appendix:repro}.
    Previous work has indicated that simulations of the somewhat-suppressed TFBI tend to provide reasonably accurate, though perhaps slightly underestimated, estimates of turbulent heating \citep{Evans2025}.
  
    The eighth and ninth columns of Table~\ref{tab:tfbi_suite_params} indicate the spatial and temporal extent of each 2D simulation.
    These extents correspond in each case to at least 5 times the predicted dominant TFBI wavelength in the linear regime and 40 times the predicted TFBI growth rate, though the extents were further increased in some cases to better resolve the turbulence.
    In some simulations --- such as all runs with $\kappa_{e}\geq{10}^{1.5}$, where the corresponding heavy ions' $\kappa_{i}$ become larger than 1 --- we further increased extents because the TFBI theory solver predicted peak growth rates at wavelengths near ion collisional mean free paths, where kinetic effects cause significant damping in EPPIC; in these cases the TFBI theory solver overestimates growth rates and underestimates the dominant wavelength because it does not account for kinetic effects.
    The relevant spatial and temporal scales across the broad range of parameters in the TFBI simulation suite span more than three orders of magnitude, with box widths ranging from 0.063 to 80.1\uu{m} and simulation durations ranging from 0.16 to 425\uu{ms}.
  
    The tenth column of Table~\ref{tab:tfbi_suite_params} indicates the magnetic field strength ($B$) in Gauss.
    The values of $B$ also span more than three orders of magnitude across the TFBI simulation suite, ranging from 0.24 to 691\uu{G}.
  
    The next column provides the ratio $n\subII{H}/n_{e}$, which varies from 0.5\% to 70.4\%.
    Although large-scale chromospheric models contain many regions with $n\subII{H}/n_{e}>95\%$, including more \iII{H} instead of heavy ions tends to suppress predicted TFBI growth \citep{Dimant2023, Evans2023, Evans2026a}.
    We considered simulating multiple regions with $n\subII{H}/n_{e}>71\%$, however these regions tend to have relatively small predicted growth rates, peaking at wavelengths more than a few hundred times larger than Debye lengths (even across a wide range of options for $n\textsub{mul}$), making them extremely computationally expensive to simulate with EPPIC.
    
    The second-to-last column indicates the zeroth-order neutral heating rate, as per equation~\eqref{eq:dTndt0}.
    Within the TFBI simulation suite, $(\inlinePdtime{T_{n}})\oZ$ varies across a broad range of values spanning more than six orders of magnitude, from $0.00165$ to $2850\uu{K/s}$.
    Rates more than a few K/s may be significant in the chromosphere.
    Still, the TFBI theory assumption of nearly-constant neutral temperature remains valid for all simulations here, as simulation duration multiplied by $(\inlinePdtime{T_{n}})\oZ$ implies a corresponding change in $T_{n}$ of 0.45\uu{K} or less.
  
    The last column indicates the ratio between electric field strength and threshold electric field strength required to drive the TFBI, ranging from $1.70$ to $6.46$.
    The threshold field in each simulation was computed by applying the TFBI theory solver (with default $\K$ limits, as described in \citet{Evans2026a}) to simulation inputs, but varying $E$ to test $E/B$~speeds: $10^{3.00},10^{3.01}, 10^{3.02},...,10^{4.00}\uu{m/s}$.
    Multiplying $B$ by the smallest $E/B$~speed having any positive TFBI growth yields the threshold field, $E\textsub{thr}$.
  
    We also simulated 10 additional points which failed to exhibit TFBI growth, despite the TFBI theory solver predicting growth.
    These 10 points are excluded from Table~\ref{tab:tfbi_suite_params} and the TFBI simulation suite.
    They all have the smallest value of $E/B$~speed corresponding to growth anywhere in the 5D grid, $E/B=10^{3.5}\uu{m/s}$; PIC particle noise might be sufficient to suppress the instability in these cases with relatively small drivers.
    Appendix~\ref{sec:appendix:hybrid_tfbi} discusses these points in more detail, and presents a newly-developed hybrid EPPIC simulation of one such point that actually does exhibit instability growth.

%%% ---------------- RESULTS ---------------- %%%
\section{Results}
  \label{sec:results}
  
  This section quantifies the role of the TFBI across the chromosphere by analyzing the full suite of TFBI simulations, as follows.
  Section~\ref{sec:results:typical_tfbi} analyzes a typical simulation from the TFBI simulation suite,
    while Appendices~\ref{sec:appendix:highlight_standing_wave}, \ref{sec:appendix:highlight_cooling}, and \ref{sec:appendix:highlight_soliton} highlight three simulations which show atypical behavior.
  Section~\ref{sec:results:tfbi_effects_across_chromo} quantifies the role of the TFBI across the chromosphere, by analyzing the full suite of TFBI simulations.
  
  \subsection{Typical TFBI Simulation Example}
    \label{sec:results:typical_tfbi}
  
    This section analyzes one simulation from the TFBI simulation suite, the simulation labeled (3,1,4,3,6)D in Table~\ref{tab:tfbi_suite_params}, which serves as a typical example.
    %All simulations in the suite exhibit density perturbations which grow exponentially, leading to a turbulent regime.
    Like the simulation analyzed here, most simulations from the suite show turbulence-driven increases of
      %the relevant scale sizes of perturbations,
      all charged species' average temperatures, 
      turbulent motions (density-weighted standard deviation of speeds),
      mean neutral heating rates inferred from equation~\eqref{eq:dTndt},
      and average current density parallel to the driving field (Pederson component),
      while the perpendicular (Hall) component of average current density decreases instead.
  
    %% (3,1,4,3,6)D megaplot & discussion %%
      \begin{figure}[htbp]
        \centering
        \includegraphics[width=\textwidth]{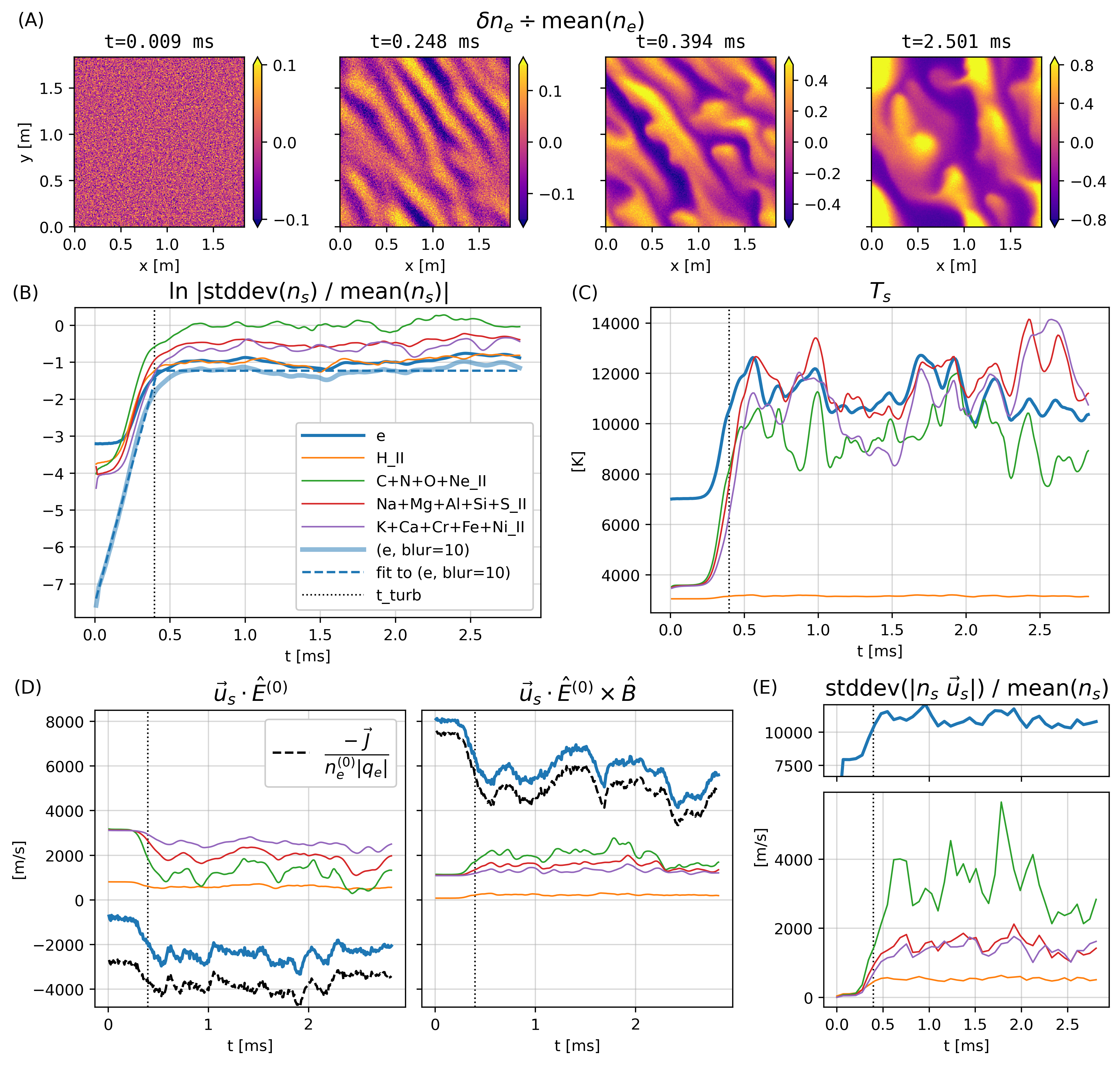}
        \caption{
          Density perturbations, temperatures, and velocities from a typical TFBI suite simulation.
          Panels~A (top four plots) show electron density perturbations relative to the mean, across the 2D simulation box at four snapshots in time.
          The remaining panels show values versus time for
            electrons (blue),
            \iII{H} (orange),
            \iII{C+N+O+Ne} (green),
            \iII{Na+Mg+Al+Si+S} (red),
            and \iII{K+Ca+Cr+Fe+Ni} (purple).
          Panel~B shows the log of density perturbation strengths (ratio between the standard deviation and the mean), and includes two additional lines:
            (thick light blue) filtered $n_{e}$, with 2D gaussian filter of width $\sigma{=}10$~grid cells;
            and (dashed blue) piecewise linear fit to the filtered $n_{e}$ values.
          The breakpoint of the linear fit determines $t\textsub{turb}$, indicated by vertical dotted black lines in Panels~B through E.
          Panels~C, D, and E show mean temperatures, velocity components, and turbulent motions, respectively.
          Panels~D also show the normalized current, $-\vec{J}/(n_{e}\oZ|q_{e}|)$ (black dashed lines).
        }
        \label{fig:typical_sim_megaplot}
      \end{figure}
  
      Figure~\ref{fig:typical_sim_megaplot} shows the density perturbations, temperatures, and velocities from simulation (3,1,4,3,6)D.
      Panels~A shows electron density perturbations relative to the mean ($\delta{n}_{e}/\text{mean}(n_{e})$, where $\delta{n}_{e}=n_{e}-\text{mean}(n_{e})$) at four snapshots in time.
        At $t{=}0.009\uu{ms}$, close to the start of the simulation, PIC particle noise dominates as no structures have formed yet.
        By $t{=}0.248\uu{ms}$, clear wave structures have developed and grow exponentially, with roughly 7 waves diagonally across the box,
          in agreement with linear TFBI theory's prediction of 5 wavelengths per simulation box side length, or $5\sqrt{2}\approx7.07$ waves across the diagonal, for waves closer to 45\degree.
        Most simulations in the TFBI simulation suite have a similar number of waves across the box during the linear regime.
        By $t{=}0.394\uu{ms}$ (the closest saved snapshot time to $t\textsub{turb}{=}0.397\uu{ms}$, as defined in the next paragraph), the linear wave structures have started to break down, as turbulence develops when density perturbations become significant relative to the background density.
        At $t{=}2.501\uu{ms}$, the turbulent regime is well underway, and the scale sizes of features in the box have increased, with now only a few large-scale structures fitting across the box.
      While the simulation is not large enough to properly resolve the turbulent regime by allowing many features across the box, prior work demonstrated this lack of resolution will likely lead to slightly underestimated but still reasonably accurate measurements of turbulent effects \citep{Evans2025}.
  
      Panel~B of Figure~\ref{fig:typical_sim_megaplot} shows the log of density perturbation strengths (ratio between the standard deviation and the mean) for all charged species.
      From $t{\approx}0.2$~to~$0.4\uu{ms}$, the slopes are roughly constant and positive, corresponding to exponential growth during the linear regime of the TFBI.
      After $t{\approx}0.4\uu{ms}$, the slopes decrease then fluctuate around 0, as the perturbation strengths reach saturation.
      This common behavior motivates a fitting procedure to determine the time of turbulent onset ($t\textsub{turb}$): fit a piecewise linear function with two pieces, the first with a positive slope and the second with a perfectly flat slope, to the log of perturbation strengths.
      Then, define $t\textsub{turb}$ as the breakpoint of this function, where the slope changes.
      To reduce the impact of particle noise on the fit, we first filter the $n_{e}$ perturbation strength by applying a 2D gaussian filter with width $\sigma{=}10$~grid cells.
      This produces the thick light blue line in Panel~B, which has far less variation in slope before $t{=}0.4\uu{ms}$ than the unfiltered values.
      The dashed blue line in Panel~B indicates the resulting piecewise linear fit, which has breakpoint $t\textsub{turb}=0.3968\pm0.0055\uu{ms}$, with error representing the standard deviation from the fitting routine (non-linear least squares via \code{scipy.optimize.curve_fit}).
      %We leave a detailed investigation of error introduced by choice of $t\textsub{turb}$ to future work, and do not represent this error on the plot nor propagate it forward to other results.
      Averaging all values after $t\textsub{turb}$ provides one measurement of turbulent properties.
      For example, taking the mean and standard deviation across all times after $t\textsub{turb}$ yields the (unfiltered) electron saturation level: $\ln|\text{stddev}(n_{e})/\text{mean}(n_{e})|=-0.98\pm0.12$.
  
      Panel~C of Figure~\ref{fig:typical_sim_megaplot} shows the average temperatures of each species over time.
      These averages correspond to the second moments of the distribution function of all particles, treating the entire simulation box as a single plasma element.
      The second moments provide values along each direction ($T_{x}$, $T_{y}$, and $T_{z}$), which we combine into a single temperature via $T=\sqrt{\frac{1}{2} \parens{T_{x}^2+T_{y}^2}}$, ignoring $T_{z}$ because out-of-plane dynamics are not properly resolved in the 2D simulation.
      This method for computing temperature follows Appendix~C of \citet{Evans2025}.
      Temperatures initially correspond closely to their equilibrium values from equation~\eqref{eq:T0}, but increase noticeably due to turbulence, then fluctuate.
      For example, initially the electron temperature is roughly $T_{e}\oZ\approx7000\uu{K}$, but it increases to $T_{e}\textsup{turb}=11160\pm710$~K after $t\textsub{turb}$, with error representing one standard deviation.
      To compute turbulent heating, defined in this work as turbulent temperature minus background temperature before any linear growth, we first take a more precise measurement of the background value by averaging $T_{e}$ between $0.05t\textsub{turb}<t<0.20t\textsub{turb}$, finding $T_{e}\textsup{bg}=7021.6\pm2.4\uu{K}$.
      The $0.05 t\textsub{turb}$ lower bound avoids startup noise, while the $0.20 t\textsub{turb}$ upper bound avoids times with TFBI waves and growth.
      Subtracting these values (with errors adding in quadrature) yields turbulent heating of electrons
        $\Delta{T}_{e}\textsup{turb}=T_{e}\textsup{turb}-T_{e}\textsup{bg}=4140\pm710\uu{K}$.
      Performing similar computations for the other species yields
        $\Delta{T}\subII{H}\textsup{turb}=114\pm19\uu{K}$,
        $\Delta{T}\subII{C+N+O+Ne}\textsup{turb}=5920\pm950\uu{K}$,
        $\Delta{T}\subII{Na+Mg+Al+Si+S}\textsup{turb}=8140\pm1020\uu{K}$, and
        $\Delta{T}\subII{K+Ca+Cr+Fe+Ni}\textsup{turb}=7430\pm1490\uu{K}$.
      Throughout the remainder of this work, we use an analogous process to measure the effects of turbulence across other quantities and simulations.
  
      Panels~D of Figure~\ref{fig:typical_sim_megaplot} show the average velocities of each species over time.
      These averages correspond to the first moments of the distribution function of all particles, treating the entire simulation box as a single plasma element, which yields a single $u_{s,x}$ and $u_{s,y}$ for each species at each snapshot in time.
      % The $|u_{s,z}|$ contributions to speed are always small, as $|u_{s,z}| < 0.061 \max{\parens{|u_{s,x}|, |u_{s,y}|}}$ across all times in this simulation.
      The left plot shows the Pederson component of velocities (sometimes also referred to as the ambipolar component), $\vec{u}_{s}\cdot\hat{E}\oZ=u_{s,x}$, while the right plot shows the Hall component, $\vec{u}_{s}\cdot\hat{E}\oZ\times\hat{B}=-u_{s,y}$.
      Both plots also show the corresponding components of the average current density (charge density-weighted sum of average velocities) normalized by the mean electron charge density ($n_{e}\oZ{q}_{e}=-1.58\times10^{-7}~C/m^3$, corresponding to $n_{e}\oZ=9.88\times10^{11}\uue{m}{-3}$) in order to display them on the same scale.
      The collective turbulence-driven changes to charged species' velocities cause the Pederson component of current density to increase by $(|\vec{J}\cdot\hat{E}\oZ|\textsup{turb}/|\vec{J}\cdot\hat{E}\oZ|\textsup{bg})-1=(37.0\pm13.1)\%$.
      The stronger current parallel to $\hat{E}$ will work to short out the driving electric field, consistent with \citet{Oppenheim1997}, and may also contribute to magnetic reconnection in 3D \citep[see, e.g.,][]{Schindler1988}.
      Meanwhile, turbulence causes the Hall component of current to instead decrease in magnitude by $(34.8\pm8.2)\%$.
  
      Turbulence also notably separates charged species' velocites from each other, even though their initial zeroth-order values are nearly equivalent.
      For example, the Pederson component decreases much more for \iII{C+N+O+Ne} than for \iII{K+Ca+Cr+Fe+Ni}.
      This difference might lead to chemical fractionation \citep{Testa2010, Testa2015} in the chromosphere, providing an alternative mechanism to previous works such as \citet{Laming2017, MartinezSykora2023,Wargnier2023}.
  
      Finally, Panels~E of Figure~\ref{fig:typical_sim_megaplot} show turbulent motions.
      At each point in time, we compute the amount turbulent motion as the standard deviation of $n_{s}|\vec{u}_{s}|$, then divide by $\text{mean}(n_{s})$ ($=n_{s}\oZ$).
      This quantity measures the degree to which fluid speeds vary across space, unlike the average temperatures and velocities from Panels C and D which instead ignore spatial variations by treating the entire box as a single plasma element.
      Here, fluid speed is computed in each grid cell as the magnitude of fluid velocity, which comes from the first moments of the distribution function of particles within that cell, only.
      Multiplying by densities before taking the standard deviation avoids systematically under-weighting contributions from denser regions.
      %The density weighting of cells before taking the standard deviation is comparable to the averaging performed when computing temperatures and mean velocities, as taking moments of the distribution weights all particles evenly, and grid cells with more particles have proportionally larger densities.
       
      In the turbulent regime, we find turbulent motions equal to:
        $10960\pm450$~m/s for electrons,
        $538\pm42$~m/s for \iII{H},
        $3170\pm840$~m/s for \iII{C+N+O+Ne},
        $1500\pm250$~m/s for \iII{Na+Mg+Al+Si+S}, and
        $1380\pm220$~m/s for \iII{K+Ca+Cr+Fe+Ni}.
      Within any region of space containing no variation in fluid velocities, this measure of turbulent motion would be zero.
      However, due to particle noise levels in EPPIC, the simulation shows nonzero turbulent motion even in the background conditions before any instability develops:
        7900\uu{m/s} for electrons,
        103\uu{m/s} for \iII{H},
        75\uu{m/s} for \iII{C+N+O+Ne},
        55\uu{m/s} for \iII{Na+Mg+Al+Si+S}, and
        50\uu{m/s} for \iII{K+Ca+Cr+Fe+Ni}.
      Due to subcycling EPPIC flux outputs, we do not have enough points between $0.05t\textsub{turb}<t<0.20t\textsub{turb}$ to compute standard deviations for these values; this subcycling also causes the lower temporal resolution in Panels~E.
      Compared to turbulent motions in the turbulent regime, these ``background turbulent motions'' are small enough to neglect for ions, but significant for electrons.
      TFBI-driven turbulent motion may contribute to the nonthermal line broadening seen in observations.
      Values from this simulation, of a few km/s, are consistent with
        nonthermal velocities observed in the optically thin \iI{O} line \citep{Carlsson2023}, and
        values of the microturbulence parameter inferred from observations \citep[see e.g.,][]{daSilvaSantos2020}.

    %% (3,1,4,3,6)D dTndt %%
      \begin{figure}[htbp]
        \centering
        \includegraphics[width=\textwidth]{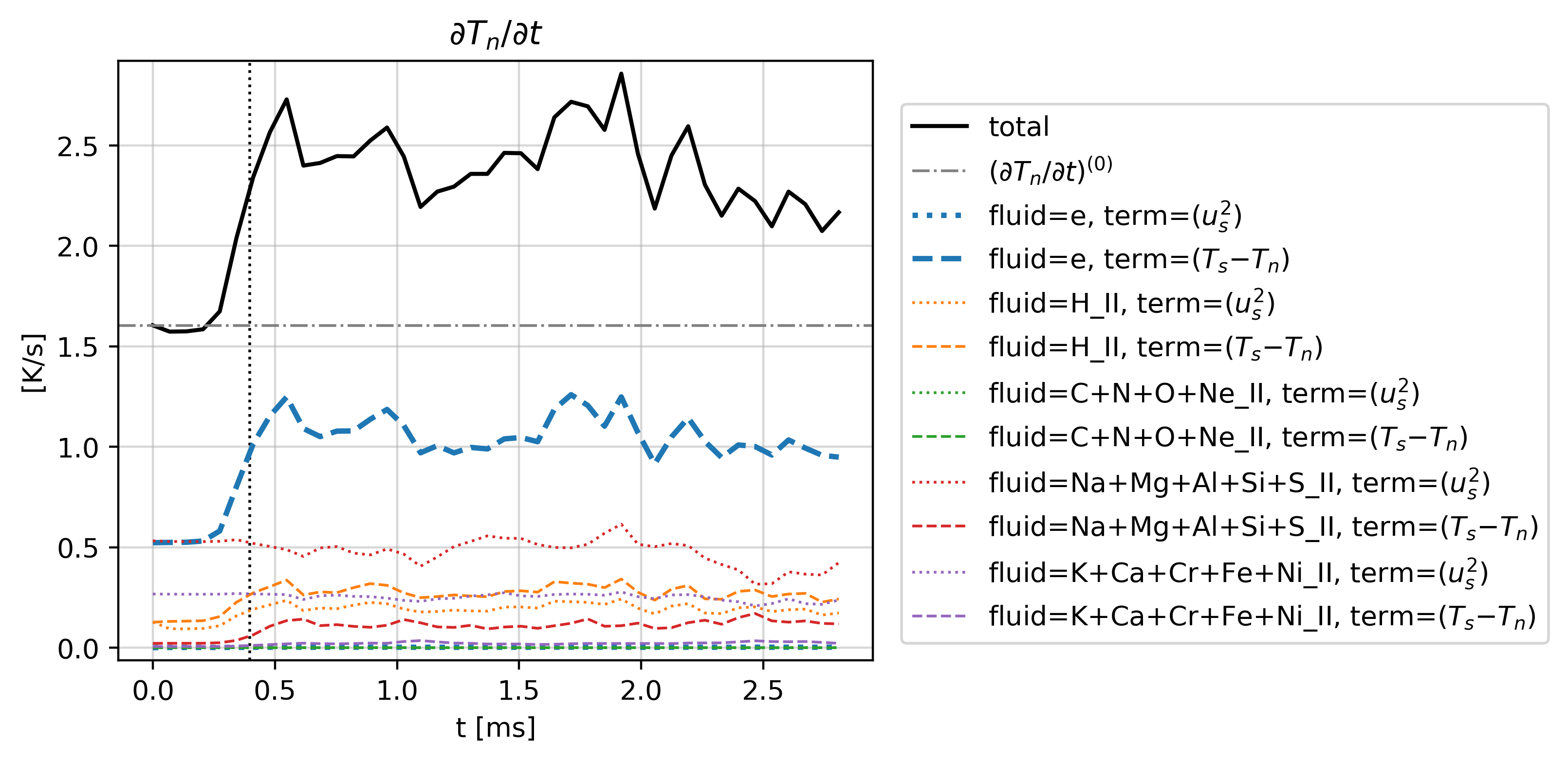}
        \caption{
          Neutral heating rate ($\inlinePdtime{T_{n}}$) versus time, for the typical TFBI suite simulation shown in Figure~\ref{fig:typical_sim_megaplot}.
          The plot includes the total neutral heating rate (solid black line) from equation~\eqref{eq:dTndt},
          as well as a breakdown of the contributions from the $|\vec{u}_{s}|^2$ (dashed lines) and $(T_{s}-T_{n})$ (dotted lines) terms, from each species (using the same color scheme as in Figure~\ref{fig:typical_sim_megaplot}).
          The horizontal dot-dashed gray line indicates the zeroth-order neutral heating rate, from equation~\eqref{eq:dTndt0},
          and the vertical dotted line indicates $t\textsub{turb}$.
        }
        \label{fig:typical_sim_dTndt}
      \end{figure}
  
      Figure~\ref{fig:typical_sim_dTndt} shows neutral heating rates versus time, computed via equation~\eqref{eq:dTndt}.
      Although the scale here shows only a few K/s, such heating may be relevant to the chromosphere where dynamical timescales can be hundreds of seconds.
      Additionally, while such increases might not seem as drastic as the thousands-of-Kelvin increases to charged species' temperatures, neutral heating rates give a much clearer picture of TFBI-driven energy transfer because the TFBI occurs in weakly ionized plasmas where the neutral density is orders of magnitude larger than charged species' densities.
      Evaluating equation~\eqref{eq:dTndt0} using this simulation's input parameters implies a zeroth-order neutral heating rate of 1.60~K/s.
      Meanwhile, the total neutral heating rate in the turbulent regime is $2.41\pm0.19\uu{K/s}$, which is larger by $(50.\pm12)\%$.
      
      The most significant contribution to $\inlinePdtime{T_{n}}$, as shown in Figure~\ref{fig:typical_sim_dTndt}, is the electron $(T_{s}-T_{n})$ term, representing collision-mediated thermalization with the heated electrons.
      The next-largest contribution is the $|\vec{u}_{s}|^2$ term for \iII{Na+Mg+Al+Si+S}, representing frictional heating due to collisions with this heavy ion species.
      Frictional heating and thermalization with \iII{H}, \iII{Na+Mg+Al+Si+S}, and \iII{K+Ca+Cr+Fe+Ni} all play a role, contributing at least 1\% to the total.
      Frictional heating with electrons never amounts to more than 0.2\% of the total, which is not too surprising as the frictional heating term scales with charged species mass, unlike the thermalization term.
      The \iII{C+N+O+Ne} species contributes less than $10^{-6}$ of the total, largely owing to its tiny number density in this simulation, where $n_{i}\oZ/n_{e}\oZ=0.500$, $5.18\times10^{-7}$, $0.334$, and $0.166$ for $i{=}$\iII{H}, \iII{C+N+O+Ne}, \iII{Na+Mg+Al+Si+S}, and \iII{K+Ca+Cr+Fe+Ni}, respectively.
      Across other TFBI suite simulations, the relative contributions to $\inlinePdtime{T_{n}}$ vary somewhat, though it is common for the dominant contributor to be thermalization with electrons, followed by frictional heating with one of the heavy ion species.

  \subsection{Effects of the TFBI Across the Chromosphere}
    \label{sec:results:tfbi_effects_across_chromo}
  
    This section quantifies TFBI effects across the chromosphere, by compiling results of analyses similar to those of Section~\ref{sec:results:typical_tfbi}, but repeated across the entire TFBI simulation suite.
  
    \subsubsection{TFBI-driven Turbulent Heating}
    %% Turbulent heating vs relative surplus driving field %%
      \begin{figure}[htbp]
        \centering
        \includegraphics[width=\textwidth]{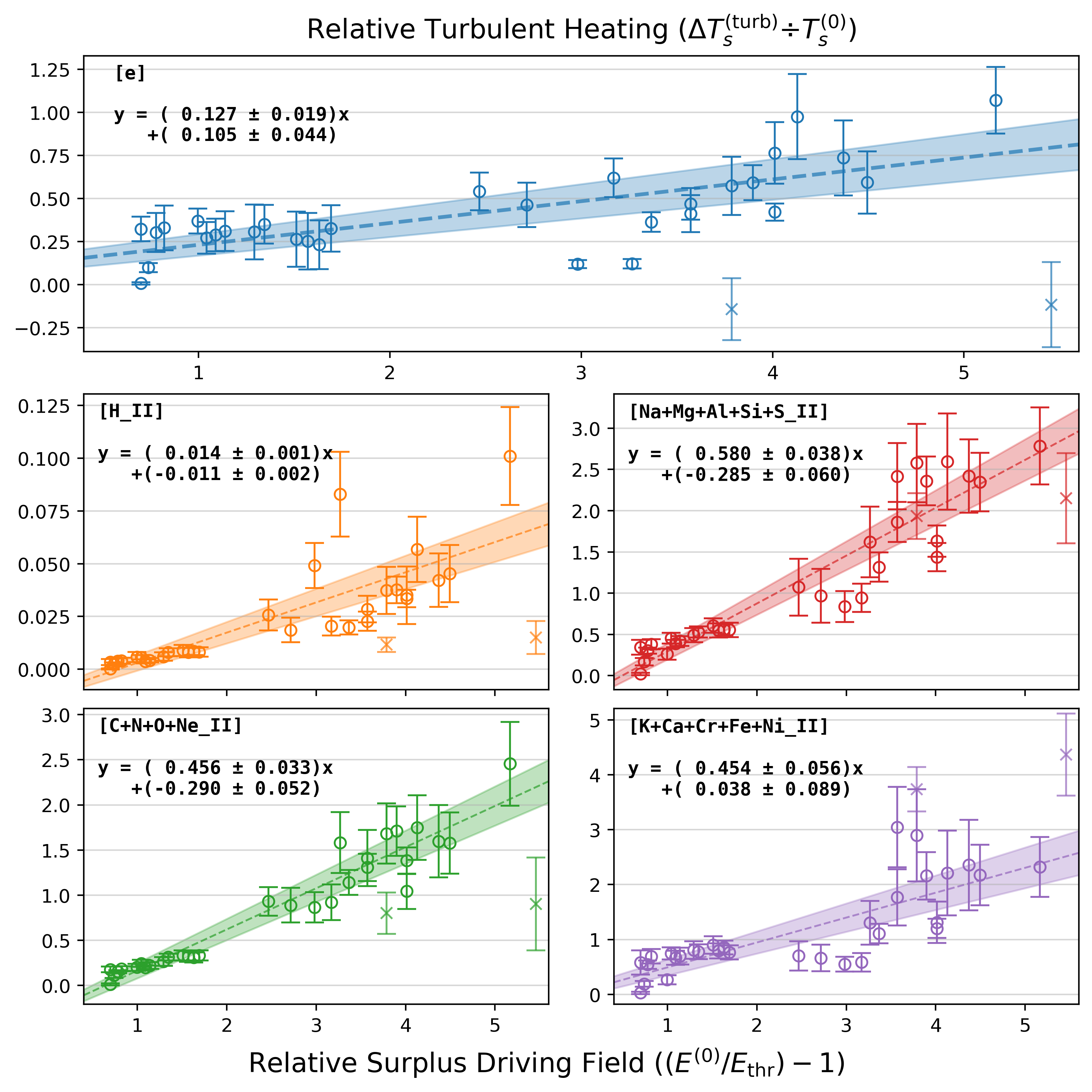}
        \caption{
          Relative turbulent heating of each charged species ($\Delta T_{s}\textsup{turb}/T_{s}\oZ$) versus relative surplus driving field ($(E\oZ/E\textsub{thr})-1$), across the TFBI simulation suite.
          Each point represents the value computed from a single simulation, with error bars indicating one standard deviation, analogously to the computations from Section~\ref{sec:results:typical_tfbi}.
          Text in the top left corner of each panel indicates the plotted species, as well as the slope and intercept of the line of best fit (plotted as dashed lines), with errors indicating one standard deviation (plotted as colored regions).
          Fits exclude the two TFBI suite simulations with $E\oZ/B=10^{4.5}\uu{m/s}$, plotted here with ``x'' markers.
        }
        \label{fig:heating_vs_Erel}
      \end{figure}
  
      Figure~\ref{fig:heating_vs_Erel} shows relative turbulent heating of each species versus relative surplus driving field, across the TFBI simulation suite.
      Each point represents the value ($\Delta T_{s}\textsup{turb}/T_{s}\oZ$) computed from a single simulation for a given charged species~$s$, following the same process as in Section~\ref{sec:results:typical_tfbi}.
      For example, the point (3.9, $0.59\pm0.10$) in the top panel (electrons) comes from simulation (3,1,4,3,6)D, which has
        ${(E\oZ/E\textsub{thr})-1}=3.9$ as per Table~\ref{tab:tfbi_suite_params},
        $\Delta{T}_{e}\textsup{turb}=4140\pm710\uu{K}$ as per Section~\ref{sec:results:typical_tfbi},
        and $T_{e}\oZ=7009$~K as per equation~\eqref{eq:T0},
        giving $\Delta{T}_{e}\textsup{turb}/T_{e}\oZ=0.59\pm0.10$.
      The plots show clear trends, with relative turbulent heating increasing proportionally to relative surplus driving field, aside from a few outliers.
  
      To quantify the trends, we use a sampling process to compute the line of best fit for each species.
      We generate one set of samples by picking one value per point, randomly selected based on the gaussian distribution corresponding to that point's value and error bar.
      For a given set of samples, we compute the best fit line using linear least squares regression.
      To get the overall line of best fit, we generate 10,000 such sets of samples, compute the best fit line for each, then take the mean slope and intercept across all 10,000 fits.
      The top left corner of each panel of Figure~\ref{fig:heating_vs_Erel} reports the overall result for each species, with errors indicating standard deviation of slope and intercept across all 10,000 fits.
  
      These fits, and all other fits following the same procedure throughout this work, exclude the two simulations (plotted with ``x'' markers) with the largest $E/B$~speed ($E\oZ/B=10^{4.5}\uu{m/s}$).
      Both of these simulations have $T_{e}\oZ>40000\uu{K}$ (compare with all other TFBI suite simulations, which have $T_{e}\oZ<9500\uu{K}$), which might introduce significant radiative effects if present in the actual chromosphere.
      However, radiation physics are currently neglected in the TFBI theory and simulations.
      Additionally, these two simulations both exhibit turbulent \emph{cooling} of electrons, an unexpected phenomenon which is explored further in Appendix~\ref{sec:appendix:highlight_cooling}.
  
      The resulting fits quantify how much turbulent heating can be expected from the TFBI, across a broad range of parameters.
      For example, TFBI driven with $E\oZ/E\textsub{thr} = 2$ can be expected to increase
        electron temperatures by $(23.1\pm6.2)\%$ compared to equilibrium values,
        with analogous increases for \iII{H}, \iII{C+N+O+Ne}, \iII{Na+Mg+Al+Si+S}, and \iII{K+Ca+Cr+Fe+Ni} of
        $(0.31\pm0.37)\%$, $(16.6\pm8.5)\%$, $(29.5\pm9.8)\%$, and $(49.2\pm14.5)\%$, respectively.
      Relative turbulent heating is most significant for the heavier ions, though at small enough values of $E\oZ/E\textsub{thr}$, the relative turbulent heating of electrons is larger than that of \iII{C+N+O+Ne}.
      Meanwhile, the relative turbulent heating of \iII{H} is almost negligible.
      Even at $E\oZ/E\textsub{thr}=5$, the relative turbulent heating of \iII{H} is only $(4.60\pm0.79)\%$.
      Note that this study likely somewhat underestimates turbulent heating (and other turbulent effects), for reasons further discussed in Section~\ref{sec:methods:tfbi_suite_params} and Section~\ref{sec:discussion}.
  
      In addition to the points excluded from fitting procedures, there are a few more outliers in the top panel (for electrons) of Figure~\ref{fig:heating_vs_Erel}.
      These outliers appear at points
        (0.7, $0.0072\pm0.0066$),
        (3.0, $0.12\pm0.02$), and
        (3.3, $0.12\pm0.03$),
      which correspond to simulations labeled (2,0,3,3,4)A, (3,1,4,4,6)E, and (3,0,3,4,6)C, respectively.
      Simulation (3,1,4,4,6)E develops a long-lasting, slowly-propagating structure within the ion densities (but not the electron density) in the turbulent regime, and is analyzed in more detail in Appendix~\ref{sec:appendix:highlight_soliton}.
      While no similar structure develops in simulation (3,0,3,4,6)C, it is the only other simulation in the TFBI suite with $|\kappa_{e}|\geq10^{1.5}$, suggesting that increasing $|\kappa_{e}|$ too much might mitigate turbulence-driven electron temperature increases.
      Simulation (2,0,3,3,4)A might deviate from the trend due to underresolved turbulence, as the simulation has only five waves across the diagonal during the linear regime (as opposed to the usual minimum box size for TFBI suite simulation of at least five waves per box edge length, i.e. $5\sqrt{2}\approx7.07$ waves across the diagonal).
      Because simulation (2,0,3,3,4)A is already one of the most computationally expensive runs in the suite, we leave a deeper investigation to future work.

    \subsubsection{TFBI-driven Turbulent Motions}
    %% Turbulent motions vs relative surplus driving field %%
      \begin{figure}[htbp]
        \centering
        \includegraphics[width=\textwidth]{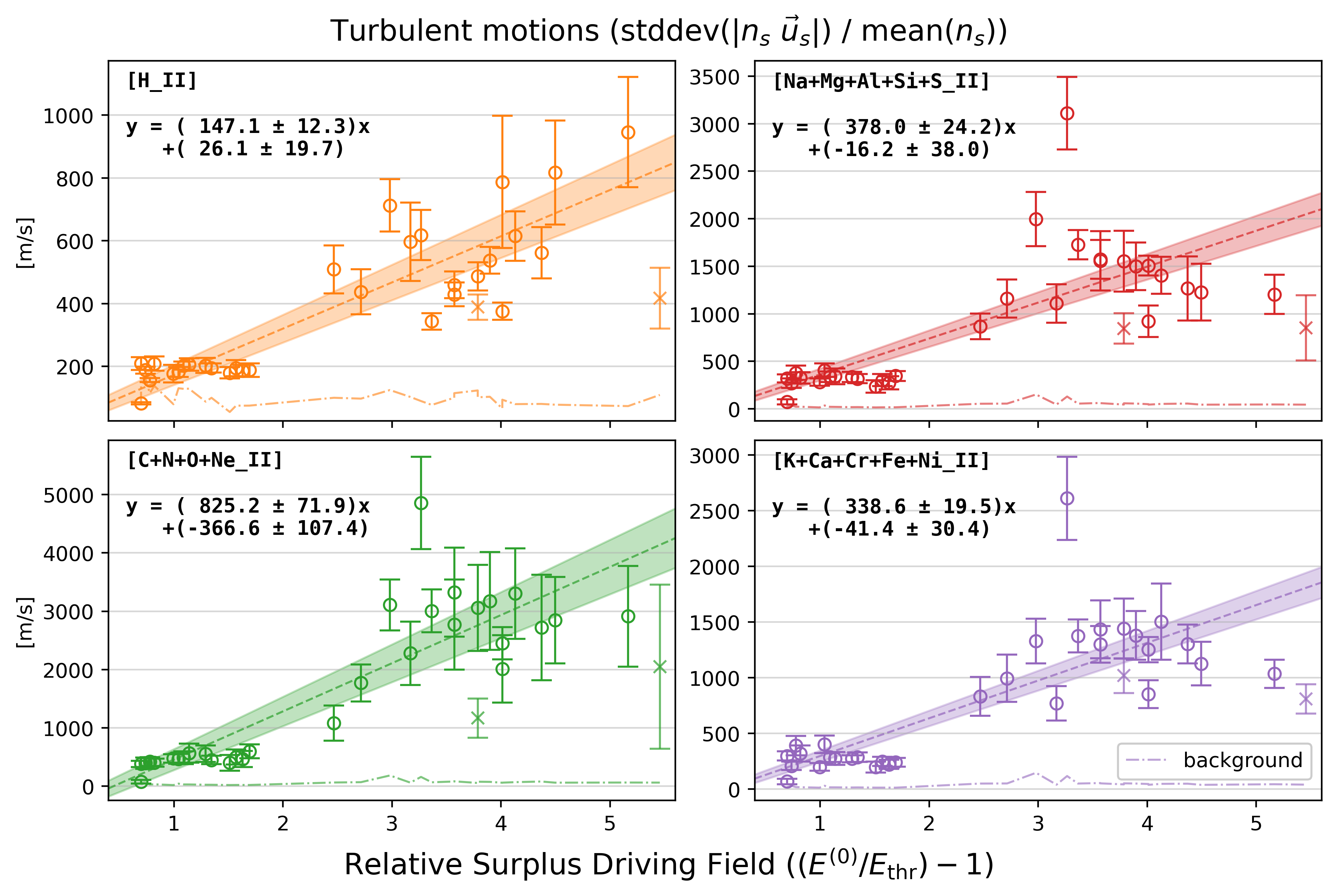}
        \caption{
          Turbulent motions of each charged species ($\text{stddev}(n_{s}|\vec{u}_{s}|)/\text{mean}(n_{s})$) versus relative surplus driving field (${(E\oZ/E\textsub{thr})-1}$), across the TFBI simulation suite.
          Dot-dashed lines near the bottom of each panel indicate ``background turbulent motions'' (due to particle noise) for each species.
          Electrons are excluded here due to significant background turbulent motions.
          Formatting is otherwise similar to Figure~\ref{fig:heating_vs_Erel}.
        }
        \label{fig:turbmotions_vs_Erel}
      \end{figure}
  
      Figure~\ref{fig:turbmotions_vs_Erel} plots turbulent motions ($\text{stddev}|n_{s} \vec{u}_{s}| / \text{mean}(n_{s})$) in the turbulent regime for each ion species across the TFBI simulation suite.
      The plot formatting and fitting procedures are similar to that of Figure~\ref{fig:heating_vs_Erel}.
      Figure~\ref{fig:turbmotions_vs_Erel} also includes dot-dashed lines to indicate the background turbulent motions, computed with the earliest available velocity data after $t=0$.
      %(TFBI suite simulations only saved velocity arrays every eighth snapshot.)
      Background turbulent motions are indicative of background noise level before the TFBI develops (e.g., due to particle noise), and depend directly on numerical parameters (e.g., number of PIC particles per cell).
      Electron background turbulent motions are always more than 55\% of turbulent motion in the turbulent regime, far too large to ignore, so we exclude electrons from this plot.
      % if reviewer asks, can add something about: it's much larger for electrons mainly due to having fewer PIC electrons than PIC ions, and subcycling. Subcycling also means that the electric field felt by the ions is less noisy than the 55\% seems to suggest.
      Meanwhile, heavy ion background turbulent motions are less than 15\% of turbulent motion in the turbulent regime in all simulations except one (simulation (2,0,3,3,4)A), and are smaller than the size of the error bars in all cases.
      For \iII{H}, background turbulent motions are less than 50\% of turbulent motion in the turbulent regime for all simulations with ${(E\oZ/E\textsub{thr})-1}>1.35$.
  
      Figure~\ref{fig:turbmotions_vs_Erel} demonstrates that turbulent motions of all ion species increase roughly proportionally to the relative surplus driving field.
      TFBI driven with $E\oZ/E\textsub{thr}=2$ can be expected to cause ion turbulent motions of up to $460\pm180\uu{m/s}$, with the largest motions in \iII{C+N+O+Ne} and lesser motions for the other ion species.
      Meanwhile, TFBI driven with $E\oZ/E\textsub{thr}=5$ can be expected to cause ion turbulent motions of roughly:
        $610\pm70$, $2930\pm400$, $1500\pm130$, and $1310\pm110\uu{m/s}$ for \iII{H}, \iII{C+N+O+Ne}, \iII{Na+Mg+Al+Si+S}, and \iII{K+Ca+Cr+Fe+Ni}, respectively.
      
      There are a few outliers in the ion turbulent motion plots.
      Simulations near ${(E\oZ/E\textsub{thr})-1}\approx1.5$ systematically fall below the trend line, hinting at the possibility that the true underlying relationship might not be linear.
      Perhaps a piecewise linear fit or a step function might be a better fit to the data, but we limit this study to linear fits to reduce the possibility of overparameterization.
      Beyond ${(E\oZ/E\textsub{thr})-1}>2.0$, most outliers fall within 2 standard deviations of the trend, though the point at ${(E\oZ/E\textsub{thr})-1}=3.27$ stands out as having much larger turbulent motions in all heavy ions.
      This point corresponds to simulation (3,0,3,4,6)C, which is also an outlier in Figure~\ref{fig:heating_vs_Erel}, as previously discussed.

    \subsubsection{TFBI-driven Changes to Average Velocities and Current}
    %% Turbulent transport: effects on average velocity %%
      % [INTENTIONALLY NO FIGURE FOR THIS QUANTITY]
      To study TFBI-driven changes to average velocities and current, we considered magnitude, direction, and individual vector components versus relative surplus driving field, finding some common behaviors but no clear linear trends.
        Across all TFBI suite simulations, turbulence causes the Hall component of average current ($\vec{J}\cdot\hat{E}\oZ\times\hat{B}$) to decrease in magnitude between 1\% and 84\%, with a mean decrease of 54\%.
        Meanwhile, the Pederson component ($\vec{J}\cdot\hat{E}\oZ$) increases in magnitude in all but three simulations.
        Across the simulations where Pederson current increases, the increases range from 4\% to 62\%, with a mean of 31\%.
        One of the simulations where it decreases instead (simulation (3,0,1,2,8)B) is also consistent with a slight increase, to within one standard deviation, as it has $(|\vec{J}\cdot\hat{E}\oZ|\textsup{turb}/|\vec{J}\cdot\hat{E}\oZ|\textsup{bg})-1=(-0.85\pm18.5)\%$.
        The other two simulations with Pederson current decreases are the outliers already plotted with ``x'' markers in previous figures (Figure~\ref{fig:heating_vs_Erel} and Figure~\ref{fig:turbmotions_vs_Erel}).
        As discussed in Section~\ref{sec:results:typical_tfbi}, the turbulence-driven increases to Pederson current which occur in most cases serve to short out the driving electric field.

    \subsubsection{TFBI-driven Neutral Heating}
    %% Turbulent neutral heating rate %%
      \begin{figure}[htbp]
        \centering
        \includegraphics[width=0.7\textwidth]{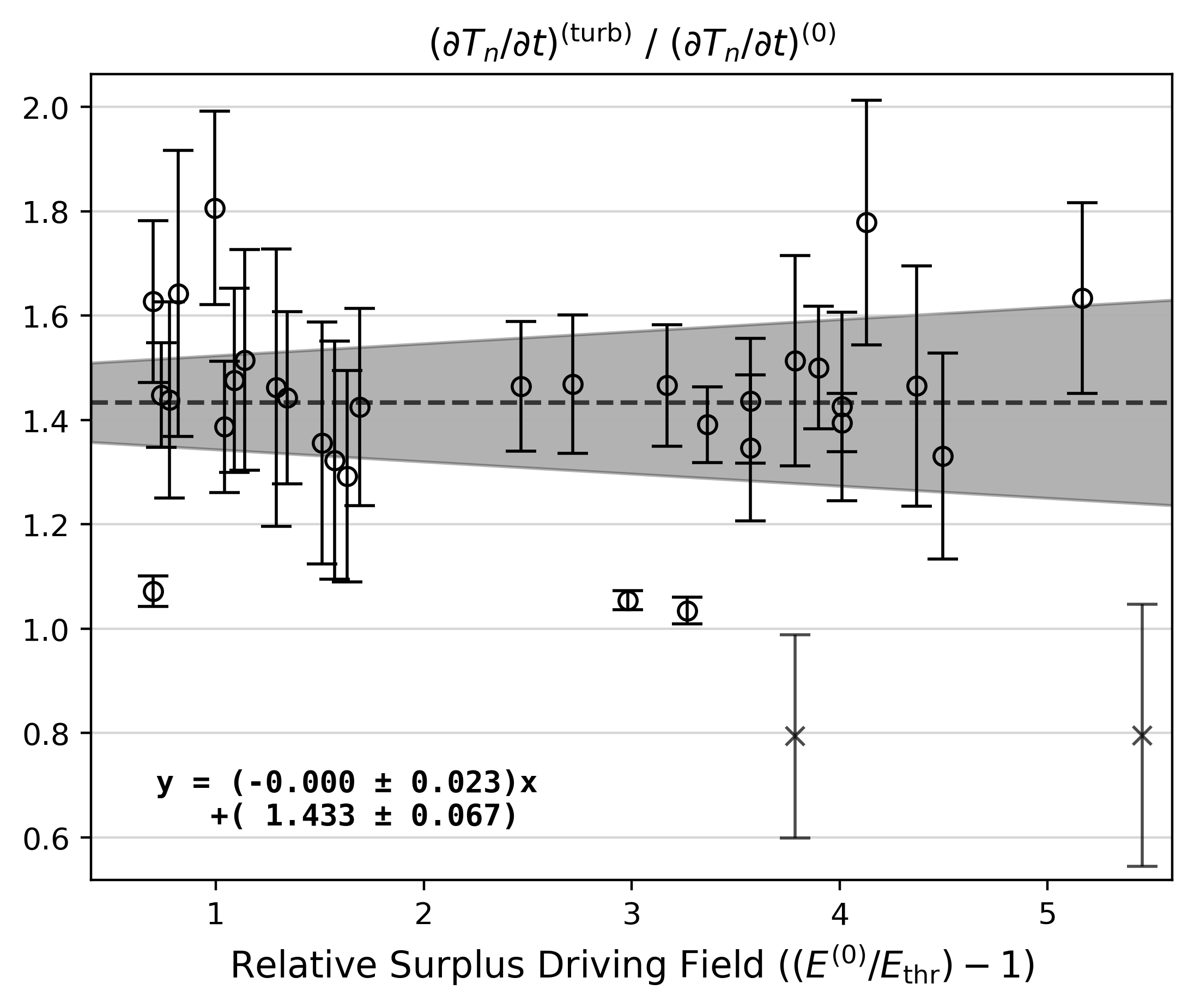}
        \caption{
          Turbulence-driven relative increase to inferred neutral heating rate from equation~\eqref{eq:dTndt} versus relative surplus driving field (${(E\oZ/E\textsub{thr})-1}$), across the TFBI simulation suite.
          Similarly to Figure~\ref{fig:heating_vs_Erel}, text indicates the slope and intercept of the line of best fit (dashed line), with errors indicating one standard deviation (shaded region), and fits exclude the same two simulations as before (plotted here with ``x'' markers).
        }
        \label{fig:dTndt_vs_Erel}
      \end{figure}
  
      Turbulence-driven changes to temperatures and velocities both contribute to inferred neutral heating rate, shown in Figure~\ref{fig:dTndt_vs_Erel}.
      The figure plots the ratio between neutral heating rates in the turbulent regime ($(\inlinePdtime{T_{n}})\textsup{turb}$; see equation~\eqref{eq:dTndt}) and zeroth-order neutral heating rates ($(\inlinePdtime{T_{n}})\oZ$; see equation~\eqref{eq:dTndt0}) versus relative surplus driving field.
      The best fit line is flat:
      \begin{equation} \label{eq:dTndt_fit}
        %% EXTRA SPACING %%
          % \frac{(\inlinePdtime{T_{n}})\textsup{turb}}{(\inlinePdtime{T_{n}})\oZ}
          %   = (0.000 \pm 0.023) \brackets{\frac{E\oZ}{E\textsub{thr}} - 1}
          %   + (1.433 \pm 0.067)
        %% APJ SPACING (remove all unnecessary internal spaces) %%
        \frac{(\inlinePdtime{T_{n}})\textsup{turb}}{(\inlinePdtime{T_{n}})\oZ}=(0.000\pm0.023)\brackets{\frac{E\oZ}{E\textsub{thr}}-1}+(1.433\pm0.067)
      \end{equation}
      This flat line indicates that on average, the TFBI increases neutral heating rates by $(43\pm7)\%$ wherever it occurs, regardless of the driving field.
      %Additionally, there is not too much spread; 70.\% (23/33) of all TFBI suite simulations fall within two standard deviations, i.e., have mean ratio $1.29 < (\inlinePdtime{T_{n}})\textsup{turb} / (\inlinePdtime{T_{n}})\oZ < 1.57$, while 85\% (28/33) have mean ratio greater than $1.29$.
      We emphasize that this relationship was derived from simulations spanning a wide range of chromospheric parameters, including more than six orders of magnitude in zeroth-order neutral heating rates, from $0.00165$ to $2850\uu{K/s}$.
  
      In addition to the points excluded from fitting procedures, there are a few outliers in Figure~\ref{fig:dTndt_vs_Erel}.
      These outliers appear at points
        (0.7, $1.072\pm0.029$),
        (3.0, $1.055\pm0.019$), and
        (3.3, $1.035\pm0.026$),
      which correspond to simulations (2,0,3,3,4)A, (3,1,4,4,6)E, and (3,0,3,4,6)C, respectively.
      These are the same three simulations which produced outliers in electron turbulent heating in Figure~\ref{fig:heating_vs_Erel}.
      Because electron temperatures tend to provide the largest contributions to inferred neutral heating rates, as shown in Figure~\ref{fig:typical_sim_dTndt} and the related discussion, it is not too surprising that these three simulations would also produce outliers here.
      Aside from these outliers, almost all points fall within one standard deviation of the line of best fit, as indicated by error bars overlapping with the shaded region in Figure~\ref{fig:dTndt_vs_Erel}.

%%% ---------------- DISCUSSION ---------------- %%%
\section{Discussion}
  \label{sec:discussion}
  
  This work quantifies the effects of the TFBI
    on turbulent heating, turbulent motion, average velocities and current density, and neutral heating rates.
    These effects can be incorporated into large-scale models wherever the TFBI is predicted to occur by modifying the corresponding physical parameters accordingly.
    Although, these results might somewhat underestimate the true effects of the TFBI, due to finite simulation size, charge separation terms introduced by artificial density scaling, and excluding 3D effects.
    Additionally, the results might be affected by selection bias in the TFBI suite.
    Despite these limitations, in this work we effectively quantify the effects of TFBI-driven turbulence across a wide range of chromospheric parameters.
  
  This work provides clear predictions about TFBI-driven turbulent heating and turbulent motions, discovering that they vary linearly with the relative surplus driving field.
    These relationships hold across most of the TFBI simulation suite, with only a few outliers.
    They may thus be used by future studies to predict TFBI physics across chromospheric models, with reasonable accuracy, without needing to perform more TFBI simulations; simply compute the relative surplus driving field from the physical parameters, then plug it into the linear relationships from Figures~\ref{fig:heating_vs_Erel} and \ref{fig:turbmotions_vs_Erel} to predict how the TFBI will affect turbulent heating and turbulent motions.
    Turbulent motions may be compared directly with nonthermal line broadening, or the microturbulence parameter inferred from inversions of observations \citep[see e.g.,][]{daSilvaSantos2020, Carlsson2023, SainzDalda2024}; here we discovered that values of up to a few~km/s might be explained by TFBI-driven ion turbulent motions.
    % Direct comparison to observation might be challenging, though, without a good way to constrain horizontal electric field in the neutral reference frame.
  
  This work also discovered that the TFBI consistently causes the magnitude of average current density components to increase parallel to $\vec{E}\oZ$, and decrease in the perpendicular direction.
    These increases in the parallel direction (Pederson/ambipolar component) correspond to current working to short out the driving electric field.
    They occur in almost all cases, though the effect varies from only a few percent up to 62\% across the TFBI suite simulations and there is no clear linear trend with relative surplus driving field.
    The decreases in the $\vec{E}\oZ\times\vec{B}$ direction (Hall component) similarly feature no clear linear trend, though they occur in all cases, varying from a few percent up to 84\%.
    These effects may cause significant changes to current densities across the chromosphere.
    One way to incorporate these physics into large-scale models could be to implement them as anomalous conductivities, analogously to what has been done for pure Farley-Buneman effects in the ionosphere \citep{Liu2016, Wiltberger2017}.
  
  Additionally, TFBI-driven turbulence separates the average velocities of heavy ions.
    Numerous studies have investigated how chemical fractionation varies as a function of first ionization potential (FIP) in the solar atmosphere \citep[see, e.g.,][and references therein]{Laming2015, Testa2015}.
    This dependence implies that chemical fractionation largely occurs in the solar chromosphere, where elements with low FIP are mostly ionized while high-FIP elements remain mostly neutral, and also suggests a connection between the fractionation mechanisms, the solar wind, and the heating of the solar atmosphere \citep{Brooks2015,Testa2023,Long2024}.
    The results of this work suggest that the TFBI may play a role, as the typical simulation example studied here demonstrates that it drives a velocity drift between \iII{C+N+O+Ne}, which is made of generally high-FIP elements, and the heavier ions \iII{Na+Mg+Al+Si+S} and \iII{K+Ca+Cr+Fe+Ni}, which are composed of mostly low-FIP elements.
    Future work might estimate the importance of this effect by quantifying such drifts across the other simulations in the TFBI simulation suite, considering the magnetic field topology wherever the TFBI is predicted to occur in chromospheric models, and considering ionization fractions for each group to predict the resulting degree of TFBI-driven chemical fractionation.
  
  We also discovered that the TFBI consistently increases neutral heating rates by $(43\pm7)\%$, on average.
    Although, results from \citet{Evans2026a} suggest that the zeroth-order neutral heating effects are not yet properly implemented in large-scale models of the Sun's atmosphere, which instead include some ad-hoc heating especially in the coldest, weakly ionized regions of the chromosphere.
    Zeroth-order neutral heating effects might have a significant impact and reduce the need for ad-hoc heating terms.
    After accounting for all relevant zeroth-order effects, the rate of neutral heating can be increased by roughly 45\% to account for TFBI contributions, everywhere with predicted TFBI growth.
    To predict TFBI growth, one can either utilize the linear TFBI solver directly, or interpolate from the 5D grid in \citet{Evans2026a}.
  
  The results from this work likely somewhat underestimate the impact of the TFBI.
    This work was limited to 2D TFBI simulations, which exclude any effects from electric fields parallel to $\vec{B}$ that might provide additional heating.
    Previous work showed the additional turbulent heating of electrons in 3D at one choice of physical parameters was roughly 10 to 15\% \citep{Evans2025}.
    However, it is possible that 3D TFBI effects in other parameter regimes might provide a more significant increase to heating.
    Additionally, the 2D simulations in this work all have finite box sizes, most of which probably underresolve the turbulent regime, with only a few structures across the box.
    \citet{Evans2025} also demonstrated that, at that same choice of parameters, the limitations of finite box size can suppress turbulent effects somewhat, though the degree to which the effects were suppressed fell within one standard deviation error bars.
    Finally, all TFBI suite simulations in this work scaled charged species' densities artificially, by some scaling factor $n\textsub{mul}$, to significantly reduce computational costs.
    As shown in \citet{Evans2025} and further discussed in \citet{Evans2026a}, artificial density scaling somewhat suppresses TFBI effects.
  
  Notably, some of the trends in this work predict negative heating or turbulent motions if extrapolated into the limit of $E\oZ$ just barely larger than $E\textsub{thr}$.
    For example, near $E\oZ/E\textsub{thr}=1$, the trend for \iII{C+N+O+Ne} in Figure~\ref{fig:turbmotions_vs_Erel} preducts turbulent motions of $-370\pm110$~m/s, even though negative turbulent motions cannot occur, by definition.
    One possible explanation is that the trend lines here have been shifted downwards, compared to what actually occurs in the chromosphere, due to the ways in which this work underestimates the effects of TFBI turbulence.
    Another possibility is that these turbulent parameters do not actually maintain a linear relationship with relative surplus driving field in the chromosphere near $E\oZ/E\textsub{thr}\rightarrow1$.
    This latter possibility seems particularly likely, as observational analysis of the pure FBI in Earth's ionosphere shows the minimum field required for heating frequently exceeds the threshold field required for instability growth \citep{ZhangVarney2024}.
    We caution that extrapolating results from this work outside the range of relative surplus driving fields tested here might lead to nonphysical results. 
  
  The results of this work might also be influenced by selection bias in deciding which simulations to include in the TFBI simulation suite.
    While the suite does simulate across a wide range of chromospheric values, with many physical parameters varying across multiple orders of magnitude, it preferentially includes points at which the TFBI was relatively inexpensive to simulate in EPPIC.
    This excludes all regions with $n\subII{H}/n_{e}>70.4\%$, which may occur throughout the chromosphere.
    It also excludes 10 points at which we ran simulations but failed to see TFBI growth, though Appendix~\ref{sec:appendix:hybrid_tfbi} demonstrates that other simulation techniques, such as a hybrid version of EPPIC, may enable future work to study the TFBI at these points.
    Some points, such as those discussed in Appendix~\ref{sec:appendix:highlight_cooling} and Appendix~\ref{sec:appendix:highlight_soliton}, appear as outliers compared to the general trends, but it is still possible that similar outliers might be more common in the chromosphere than in the TFBI simulation suite itself.

%%% ---------------- CONCLUSIONS ---------------- %%%
\section{Conclusions}
  \label{sec:conclusion}
  
  In this work, we performed a suite of 2D TFBI simulations across a wide range of chromospheric parameters.
    Turbulent heating and turbulent motions increase proportionally to relative surplus driving field (${E\oZ/E\textsub{thr}-1}$), as quantified in Figures~\ref{fig:heating_vs_Erel} and \ref{fig:turbmotions_vs_Erel}.
    Turbulence-driven neutral heating rates consistently increase by $(43\pm7)\%$, on average, with no correlation to surplus driving electric field. %, which can be incorporated into large-scale models by increasing neutral heating rates accordingly, anywhere with positive predicted TFBI growth.
    The TFBI also causes $\vec{J}\cdot\hat{E}\oZ$ to increase by up to roughly 60\%, while the perpendicular component decreases in magnitude by up to roughly 80\%.
    While this work likely somewhat underestimates TFBI-driven turbulent effects, it represents the first comprehensive study quantifying such effects across a variety of chromospheric parameters.
    %Future work can and should utilize the results of this study to incorporate the effects of the Thermal Farley-Buneman Instability across large-scale models of the solar atmosphere.
    These discoveries relate turbulent properties to values computed directly from background parameters and linear theory, thus providing a much simpler alternative to running TFBI simulations, for any future work seeking to incorporate TFBI effects across large-scale models of the chromosphere.

%%% ---------------- ACKNOWLEDGEMENTS ---------------- %%%
\acknowledgments{
  We thank Yakov Dimant for numerous elucidating conversations which contributed to our analysis and understanding of the TFBI.
  
  This work utilized the \code{PlasmaCalcs} package \citep{PlasmaCalcs}.

  The authors gratefully acknowledge support from
    NSF grants
      PHY-1903416,
      AGS-2350366, and
      AGS-2531480,
    and MASGC grant 1141760.
  J.M.-S. gratefully acknowledges support by
    NASA contracts
      NNG09FA40C (IRIS) and
      80GSFC21C0011 (MUSE),
    NASA HSR grant 80NSSC26K0018,
    and NSF grants
      AGS2532363 and
      AGS2532187.
  For providing computational resources, the authors gratefully acknowledge
    the Texas Advanced Computing Center (TACC) at the University of Texas at Austin
      through project ATM 23011,
    the Advanced Cyberinfrastructure Coordination Ecosystem: Services and Support (ACCESS) program
      through project EES240116,
    and the Pleiades cluster,
      through the computing projects s1061 and s2601 from the High-End Computing (HEC) division of NASA.

  MUSE
    is led by the Lockheed Martin Solar and Astrophysics Laboratory of Palo Alto, California. MUSE is managed by the Explorer’s Program Office of NASA’s Goddard Space Flight Center in Greenbelt, Maryland, for the Heliophysics Division of NASA’s Science Mission Directorate. Lockheed Martin Advanced Technology Center, along with partner institutions, builds the MUSE instrument and spacecraft and University of California, Berkeley provides the mission operations center. MUSE benefits from international contributions supported by the Norwegian Space Agency (NOSA), the Italian Space Agency (ASI), the German Space Agency at DLR, and from the Max Planck Institute for Solar System Research (MPS).
}

%%% ---------------- REFERENCES ---------------- %%%
\bibliographystyle{aasjournal}
\bibliography{bibfile}

@ARTICLE{Birdsall1991,
       author = {{Birdsall}, C.~K.},
        title = "{Particle-in-cell charged-particle simulations, plus Monte Carlo collisions with neutral atoms, PIC-MCC}",
      journal = {IEEE Transactions on Plasma Science},
         year = 1991,
        month = apr,
       volume = {19},
       number = {2},
        pages = {65-85},
          doi = {10.1109/27.106800},
       adsurl = {https://ui.adsabs.harvard.edu/abs/1991ITPS...19...65B}
}

@ARTICLE{Carlsson2016,
       author = {{Carlsson}, Mats and {Hansteen}, Viggo H. and {Gudiksen}, Boris V. and {Leenaarts}, Jorrit and {De Pontieu}, Bart},
        title = "{A publicly available simulation of an enhanced network region of the Sun}",
      journal = {\aap},
         year = 2016,
        month = jan,
       volume = {585},
          eid = {A4},
        pages = {A4},
          doi = {10.1051/0004-6361/201527226},
archivePrefix = {arXiv},
       eprint = {1510.07581},
 primaryClass = {astro-ph.SR},
       adsurl = {https://ui.adsabs.harvard.edu/abs/2016A&A...585A...4C}
}

@ARTICLE{Carlsson2019,
       author = {{Carlsson}, Mats and {De Pontieu}, Bart and {Hansteen}, Viggo H.},
        title = "{New View of the Solar Chromosphere}",
      journal = {\araa},
         year = 2019,
        month = aug,
       volume = {57},
        pages = {189-226},
          doi = {10.1146/annurev-astro-081817-052044},
       adsurl = {https://ui.adsabs.harvard.edu/abs/2019ARA&A..57..189C}
}

@ARTICLE{Carlsson2023,
       author = {{Carlsson}, Mats and {De Pontieu}, Bart},
        title = "{An Optically Thin View of the Solar Chromosphere from Observations of the O I 1355 {\r{A}} Spectral Line}",
      journal = {\apj},
         year = 2023,
        month = dec,
       volume = {959},
       number = {2},
          eid = {87},
        pages = {87},
          doi = {10.3847/1538-4357/acf451},
archivePrefix = {arXiv},
       eprint = {2308.14067},
 primaryClass = {astro-ph.SR},
       adsurl = {https://ui.adsabs.harvard.edu/abs/2023ApJ...959...87C}
}

@ARTICLE{Centeno2017,
       author = {{Centeno}, R. and {Blanco Rodr{\'\i}guez}, J. and {Del Toro Iniesta}, J.~C. and {Solanki}, S.~K. and {Barthol}, P. and {Gandorfer}, A. and {Gizon}, L. and {Hirzberger}, J. and {Riethm{\"u}ller}, T.~L. and {van Noort}, M. and {Orozco Su{\'a}rez}, D. and {Berkefeld}, T. and {Schmidt}, W. and {Mart{\'\i}nez Pillet}, V. and {Kn{\"o}lker}, M.},
        title = "{A Tale of Two Emergences: Sunrise II Observations of Emergence Sites in a Solar Active Region}",
      journal = {\apjs},
         year = 2017,
        month = mar,
       volume = {229},
       number = {1},
          eid = {3},
        pages = {3},
          doi = {10.3847/1538-4365/229/1/3},
archivePrefix = {arXiv},
       eprint = {1610.03531},
 primaryClass = {astro-ph.SR},
       adsurl = {https://ui.adsabs.harvard.edu/abs/2017ApJS..229....3C}
}

@ARTICLE{Chintzoglou2021,
       author = {{Chintzoglou}, Georgios and {De Pontieu}, Bart and {Mart{\'\i}nez-Sykora}, Juan and {Hansteen}, Viggo and {de la Cruz Rodr{\'\i}guez}, Jaime and {Szydlarski}, Mikolaj and {Jafarzadeh}, Shahin and {Wedemeyer}, Sven and {Bastian}, Timothy S. and {Sainz Dalda}, Alberto},
        title = "{ALMA and IRIS Observations of the Solar Chromosphere. I. An On-disk Type II Spicule}",
      journal = {\apj},
         year = 2021,
        month = jan,
       volume = {906},
       number = {2},
          eid = {82},
        pages = {82},
          doi = {10.3847/1538-4357/abc9b1},
archivePrefix = {arXiv},
       eprint = {2005.12717},
 primaryClass = {astro-ph.SR},
       adsurl = {https://ui.adsabs.harvard.edu/abs/2021ApJ...906...82C}
}

@ARTICLE{daSilvaSantos2020,
       author = {{da Silva Santos}, J.~M. and {de la Cruz Rodr{\'\i}guez}, J. and {Leenaarts}, J. and {Chintzoglou}, G. and {De Pontieu}, B. and {Wedemeyer}, S. and {Szydlarski}, M.},
        title = "{The multi-thermal chromosphere. Inversions of ALMA and IRIS data}",
      journal = {\aap},
         year = 2020,
        month = feb,
       volume = {634},
          eid = {A56},
        pages = {A56},
          doi = {10.1051/0004-6361/201937117},
archivePrefix = {arXiv},
       eprint = {1912.09886},
 primaryClass = {astro-ph.SR},
       adsurl = {https://ui.adsabs.harvard.edu/abs/2020A&A...634A..56D}
}

@ARTICLE{delaCruzRodriguez2015,
       author = {{de la Cruz Rodr{\'\i}guez}, Jaime and {Hansteen}, Viggo and {Bellot-Rubio}, Luis and {Ortiz}, Ada},
        title = "{Emergence of Granular-sized Magnetic Bubbles through the Solar Atmosphere. II. Non-LTE Chromospheric Diagnostics and Inversions}",
      journal = {\apj},
         year = 2015,
        month = sep,
       volume = {810},
       number = {2},
          eid = {145},
        pages = {145},
          doi = {10.1088/0004-637X/810/2/145},
archivePrefix = {arXiv},
       eprint = {1503.03846},
 primaryClass = {astro-ph.SR},
       adsurl = {https://ui.adsabs.harvard.edu/abs/2015ApJ...810..145D}
}

@ARTICLE{DePontieu2014,
       author = {{De Pontieu}, B. and {Title}, A.~M. and {Lemen}, J.~R. and {Kushner}, G.~D. and {Akin}, D.~J. and {Allard}, B. and {Berger}, T. and {Boerner}, P. and {Cheung}, M. and {Chou}, C. and {Drake}, J.~F. and {Duncan}, D.~W. and {Freeland}, S. and {Heyman}, G.~F. and {Hoffman}, C. and {Hurlburt}, N.~E. and {Lindgren}, R.~W. and {Mathur}, D. and {Rehse}, R. and {Sabolish}, D. and {Seguin}, R. and {Schrijver}, C.~J. and {Tarbell}, T.~D. and {W{\"u}lser}, J. -P. and {Wolfson}, C.~J. and {Yanari}, C. and {Mudge}, J. and {Nguyen-Phuc}, N. and {Timmons}, R. and {van Bezooijen}, R. and {Weingrod}, I. and {Brookner}, R. and {Butcher}, G. and {Dougherty}, B. and {Eder}, J. and {Knagenhjelm}, V. and {Larsen}, S. and {Mansir}, D. and {Phan}, L. and {Boyle}, P. and {Cheimets}, P.~N. and {DeLuca}, E.~E. and {Golub}, L. and {Gates}, R. and {Hertz}, E. and {McKillop}, S. and {Park}, S. and {Perry}, T. and {Podgorski}, W.~A. and {Reeves}, K. and {Saar}, S. and {Testa}, P. and {Tian}, H. and {Weber}, M. and {Dunn}, C. and {Eccles}, S. and {Jaeggli}, S.~A. and {Kankelborg}, C.~C. and {Mashburn}, K. and {Pust}, N. and {Springer}, L. and {Carvalho}, R. and {Kleint}, L. and {Marmie}, J. and {Mazmanian}, E. and {Pereira}, T.~M.~D. and {Sawyer}, S. and {Strong}, J. and {Worden}, S.~P. and {Carlsson}, M. and {Hansteen}, V.~H. and {Leenaarts}, J. and {Wiesmann}, M. and {Aloise}, J. and {Chu}, K. -C. and {Bush}, R.~I. and {Scherrer}, P.~H. and {Brekke}, P. and {Martinez-Sykora}, J. and {Lites}, B.~W. and {McIntosh}, S.~W. and {Uitenbroek}, H. and {Okamoto}, T.~J. and {Gummin}, M.~A. and {Auker}, G. and {Jerram}, P. and {Pool}, P. and {Waltham}, N.},
        title = "{The Interface Region Imaging Spectrograph (IRIS)}",
      journal = {\solphys},
         year = 2014,
        month = jul,
       volume = {289},
       number = {7},
        pages = {2733-2779},
          doi = {10.1007/s11207-014-0485-y},
archivePrefix = {arXiv},
       eprint = {1401.2491},
 primaryClass = {astro-ph.SR},
       adsurl = {https://ui.adsabs.harvard.edu/abs/2014SoPh..289.2733D}
}

@ARTICLE{Dimant2004,
       author = {{Dimant}, Y.~S. and {Oppenheim}, M.~M.},
        title = "{Ion thermal effects on E-region instabilities: linear theory}",
      journal = {Journal of Atmospheric and Solar-Terrestrial Physics},
         year = 2004,
        month = nov,
       volume = {66},
       number = {17},
        pages = {1639-1654},
          doi = {10.1016/j.jastp.2004.07.006},
       adsurl = {https://ui.adsabs.harvard.edu/abs/2004JASTP..66.1639D}
}

@ARTICLE{Dimant2023,
       author = {{Dimant}, Y.~S. and {Oppenheim}, M.~M. and {Evans}, S. and {Martinez-Sykora}, J.},
        title = "{Unified fluid theory of the collisional thermal Farley-Buneman instability including magnetized multi-species ions}",
      journal = {Physics of Plasmas},
         year = 2023,
        month = oct,
       volume = {30},
       number = {10},
          eid = {102101},
        pages = {102101},
          doi = {10.1063/5.0155500},
       adsurl = {https://ui.adsabs.harvard.edu/abs/2023PhPl...30j2101D}
}

@ARTICLE{Evans2023,
       author = {{Evans}, Samuel and {Oppenheim}, Meers and {Mart{\'\i}nez-Sykora}, Juan and {Dimant}, Yakov and {Xiao}, Richard},
        title = "{Multifluid Simulation of Solar Chromospheric Turbulence and Heating Due to Thermal Farley-Buneman Instability}",
      journal = {\apj},
         year = 2023,
        month = jun,
       volume = {949},
       number = {2},
          eid = {59},
        pages = {59},
          doi = {10.3847/1538-4357/acc5e5},
archivePrefix = {arXiv},
       eprint = {2211.03644},
 primaryClass = {astro-ph.SR},
       adsurl = {https://ui.adsabs.harvard.edu/abs/2023ApJ...949...59E}
}

@article{Evans2025,
       author = {Evans, Samuel and Oppenheim, Meers and {Mart{\'\i}nez-Sykora}, Juan and Dimant, Yakov},
        title = "{Multifluid and Kinetic 2D and 3D Simulations of Thermal Farley–Buneman Instability Turbulence in the Solar Chromosphere}",
      journal = {\apj},
         year = 2025,
        month = jun,
    publisher = {The American Astronomical Society},
       volume = {986},
       number = {1},
        pages = {23},
          doi = {10.3847/1538-4357/adcd70},
}

@ARTICLE{Evans2026a,
       author = {{Evans}, Samuel and {Oppenheim}, Meers and {Mart{\'\i}nez-Sykora}, Juan and {Dimant}, Yakov},
        title = "{The Unstable Chromosphere: Predicting Thermal Farley-Buneman Instability Growth Across a Broad Range of Solar Chromospheric Conditions}",
      journal = {\apj},
         year = {accepted},
          doi = {10.3847/1538-4357/ae4ddd}
}

@ARTICLE{Fontenla2009,
       author = {{Fontenla}, J.~M. and {Curdt}, W. and {Haberreiter}, M. and {Harder}, J. and {Tian}, H.},
        title = "{Semiempirical Models of the Solar Atmosphere. III. Set of Non-LTE Models for Far-Ultraviolet/Extreme-Ultraviolet Irradiance Computation}",
      journal = {\apj},
         year = 2009,
        month = dec,
       volume = {707},
       number = {1},
        pages = {482-502},
          doi = {10.1088/0004-637X/707/1/482},
       adsurl = {https://ui.adsabs.harvard.edu/abs/2009ApJ...707..482F}
}

@ARTICLE{Gustafsson1975,
       author = {{Gustafsson}, B. and {Bell}, R.~A. and {Eriksson}, K. and {Nordlund}, A.},
        title = "{A grid of model atmospheres for metal-deficient giant stars. I.}",
      journal = {\aap},
         year = 1975,
        month = sep,
       volume = {42},
        pages = {407-432},
       adsurl = {https://ui.adsabs.harvard.edu/abs/1975A&A....42..407G}
}

@ARTICLE{Hansteen2023,
       author = {{Hansteen}, Viggo H. and {Martinez-Sykora}, Juan and {Carlsson}, Mats and {De Pontieu}, Bart and {Go{\v{s}}i{\'c}}, Milan and {Bose}, Souvik},
        title = "{Numerical Simulations and Observations of Mg II in the Solar Chromosphere}",
      journal = {\apj},
         year = 2023,
        month = feb,
       volume = {944},
       number = {2},
          eid = {131},
        pages = {131},
          doi = {10.3847/1538-4357/acb33c},
archivePrefix = {arXiv},
       eprint = {2211.09277},
 primaryClass = {astro-ph.SR},
       adsurl = {https://ui.adsabs.harvard.edu/abs/2023ApJ...944..131H}
}

@ARTICLE{Leenaarts2011,
       author = {{Leenaarts}, J. and {Carlsson}, M. and {Hansteen}, V. and {Gudiksen}, B.~V.},
        title = "{On the minimum temperature of the quiet solar chromosphere}",
      journal = {\aap},
         year = 2011,
        month = jun,
       volume = {530},
          eid = {A124},
        pages = {A124},
          doi = {10.1051/0004-6361/201016392},
archivePrefix = {arXiv},
       eprint = {1104.5081},
 primaryClass = {astro-ph.SR},
       adsurl = {https://ui.adsabs.harvard.edu/abs/2011A&A...530A.124L}
}

@ARTICLE{Liu2016,
       author = {{Liu}, Jing and {Wang}, Wenbin and {Oppenheim}, Meers and {Dimant}, Yakov and {Wiltberger}, Michael and {Merkin}, Slava},
        title = "{Anomalous electron heating effects on the E region ionosphere in TIEGCM}",
      journal = {\grl},
         year = 2016,
        month = mar,
       volume = {43},
       number = {6},
        pages = {2351-2358},
          doi = {10.1002/2016GL068010},
       adsurl = {https://ui.adsabs.harvard.edu/abs/2016GeoRL..43.2351L}
}

@ARTICLE{Loukitcheva2019,
       author = {{Loukitcheva}, Maria A. and {White}, Stephen M. and {Solanki}, Sami K.},
        title = "{ALMA Detection of Dark Chromospheric Holes in the Quiet Sun}",
      journal = {\apjl},
         year = 2019,
        month = jun,
       volume = {877},
       number = {2},
          eid = {L26},
        pages = {L26},
          doi = {10.3847/2041-8213/ab2191},
archivePrefix = {arXiv},
       eprint = {1905.06763},
 primaryClass = {astro-ph.SR},
       adsurl = {https://ui.adsabs.harvard.edu/abs/2019ApJ...877L..26L}
}

@ARTICLE{MartinezSykora2020.NEI,
       author = {{Mart{\'\i}nez-Sykora}, Juan and {Leenaarts}, Jorrit and {De Pontieu}, Bart and {N{\'o}brega-Siverio}, Daniel and {Hansteen}, Viggo H. and {Carlsson}, Mats and {Szydlarski}, Mikolaj},
        title = "{Ion-neutral Interactions and Nonequilibrium Ionization in the Solar Chromosphere}",
      journal = {\apj},
         year = 2020,
        month = feb,
       volume = {889},
       number = {2},
          eid = {95},
        pages = {95},
          doi = {10.3847/1538-4357/ab643f},
archivePrefix = {arXiv},
       eprint = {1912.06682},
 primaryClass = {astro-ph.SR},
       adsurl = {https://ui.adsabs.harvard.edu/abs/2020ApJ...889...95M}
}

@ARTICLE{Mulay2021,
       author = {{Mulay}, Sargam M. and {Fletcher}, Lyndsay},
        title = "{Evidence of chromospheric molecular hydrogen emission in a solar flare observed by the IRIS satellite}",
      journal = {\mnras},
         year = 2021,
        month = jun,
       volume = {504},
       number = {2},
        pages = {2842-2852},
          doi = {10.1093/mnras/stab367},
archivePrefix = {arXiv},
       eprint = {2102.03329},
 primaryClass = {astro-ph.SR},
       adsurl = {https://ui.adsabs.harvard.edu/abs/2021MNRAS.504.2842M}
}

@ARTICLE{NobregaSiverio2020.AmbipolarBifrost,
       author = {{N{\'o}brega-Siverio}, D. and {Mart{\'\i}nez-Sykora}, J. and {Moreno-Insertis}, F. and {Carlsson}, M.},
        title = "{Ambipolar diffusion in the Bifrost code}",
      journal = {\aap},
         year = 2020,
        month = jun,
       volume = {638},
          eid = {A79},
        pages = {A79},
          doi = {10.1051/0004-6361/202037809},
archivePrefix = {arXiv},
       eprint = {2004.11927},
 primaryClass = {astro-ph.SR},
       adsurl = {https://ui.adsabs.harvard.edu/abs/2020A&A...638A..79N}
}

@ARTICLE{Ondratscheck2024,
       author = {{Ondratschek}, P. and {Przybylski}, D. and {Smitha}, H.~N. and {Cameron}, R. and {Solanki}, S.~K. and {Leenaarts}, J.},
        title = "{Mg II h\&k spectra of an enhanced network region simulated with the MURaM-ChE code: Results using 1.5D synthesis}",
      journal = {\aap},
         year = 2024,
        month = dec,
       volume = {692},
          eid = {A6},
        pages = {A6},
          doi = {10.1051/0004-6361/202450788},
archivePrefix = {arXiv},
       eprint = {2410.04594},
 primaryClass = {astro-ph.SR},
       adsurl = {https://ui.adsabs.harvard.edu/abs/2024A&A...692A...6O}
}

@ARTICLE{Oppenheim1997,
       author = {{Oppenheim}, M.},
        title = "{Evidence and effects of a wave-driven nonlinear current in the equatorial electrojet}",
      journal = {Annales Geophysicae},
         year = 1997,
        month = jul,
       volume = {15},
       number = {7},
        pages = {899-907},
          doi = {10.1007/s00585-997-0899-z},
       adsurl = {https://ui.adsabs.harvard.edu/abs/1997AnGeo..15..899O}
}

@ARTICLE{Oppenheim2004,
       author = {{Oppenheim}, M.~M. and {Dimant}, Y.~S.},
        title = "{Ion thermal effects on E-region instabilities: 2D kinetic simulations}",
      journal = {Journal of Atmospheric and Solar-Terrestrial Physics},
         year = 2004,
        month = nov,
       volume = {66},
       number = {17},
        pages = {1655-1668},
          doi = {10.1016/j.jastp.2004.07.007},
       adsurl = {https://ui.adsabs.harvard.edu/abs/2004JASTP..66.1655O}
}

@ARTICLE{Oppenheim2020,
       author = {{Oppenheim}, Meers and {Dimant}, Yakov and {Longley}, William and {Fletcher}, Alex C.},
        title = "{Newly Discovered Source of Turbulence and Heating in the Solar Chromosphere}",
      journal = {\apjl},
         year = 2020,
        month = mar,
       volume = {891},
       number = {1},
          eid = {L9},
        pages = {L9},
          doi = {10.3847/2041-8213/ab75bc},
       adsurl = {https://ui.adsabs.harvard.edu/abs/2020ApJ...891L...9O}
}

@ARTICLE{Ortiz2014,
       author = {{Ortiz}, Ada and {Bellot Rubio}, Luis R. and {Hansteen}, Viggo H. and {de la Cruz Rodr{\'\i}guez}, Jaime and {Rouppe van der Voort}, Luc},
        title = "{Emergence of Granular-sized Magnetic Bubbles through the Solar Atmosphere. I. Spectropolarimetric Observations and Simulations}",
      journal = {\apj},
         year = 2014,
        month = feb,
       volume = {781},
       number = {2},
          eid = {126},
        pages = {126},
          doi = {10.1088/0004-637X/781/2/126},
archivePrefix = {arXiv},
       eprint = {1312.5735},
 primaryClass = {astro-ph.SR},
       adsurl = {https://ui.adsabs.harvard.edu/abs/2014ApJ...781..126O}
}

@software{PlasmaCalcs,
  author       = {Evans, Samuel and
                  Koontaweepunya, Rattanakorn and
                  Green, Alexander and
                  Goodman, Tess},
  title        = {PlasmaCalcs},
  month        = jan,
  year         = 2026,
  publisher    = {Zenodo},
  version      = {2025.11.0},
  doi          = {10.5281/zenodo.18381500},
  url          = {https://doi.org/10.5281/zenodo.18381500},
}

@ARTICLE{Rosenberg1998,
       author = {{Rosenberg}, M. and {Chow}, V.~W.},
        title = "{Farley-Buneman instability in a dusty plasma}",
      journal = {\planss},
         year = 1998,
        month = jan,
       volume = {46},
       number = {1},
        pages = {103-108},
          doi = {10.1016/S0032-0633(97)00104-9},
       adsurl = {https://ui.adsabs.harvard.edu/abs/1998P&SS...46..103R}
}

@ARTICLE{SainzDalda2024,
       author = {{Sainz Dalda}, Alberto and {Agrawal}, Aaryan and {De Pontieu}, Bart and {Go{\v{s}}i{\'c}}, Milan},
        title = "{IRIS$^{2+}$: A Comprehensive Database of Stratified Thermodynamic Models in the Low Solar Atmosphere}",
      journal = {\apjs},
         year = 2024,
        month = mar,
       volume = {271},
       number = {1},
          eid = {24},
        pages = {24},
          doi = {10.3847/1538-4365/ad1e55},
archivePrefix = {arXiv},
       eprint = {2211.09103},
 primaryClass = {astro-ph.SR},
       adsurl = {https://ui.adsabs.harvard.edu/abs/2024ApJS..271...24S}
}

@ARTICLE{Schindler1988,
       author = {{Schindler}, K. and {Hesse}, M. and {Birn}, J.},
        title = "{General magnetic reconnection, parallel electric fields, and helicity}",
      journal = {\jgr},
         year = 1988,
        month = jun,
       volume = {93},
       number = {A6},
        pages = {5547-5557},
          doi = {10.1029/JA093iA06p05547},
       adsurl = {https://ui.adsabs.harvard.edu/abs/1988JGR....93.5547S}
}

@ARTICLE{Song2023,
       author = {{Song}, Yongliang and {Bai}, Xianyong and {Yang}, Xu and {Cao}, Wenda and {Uitenbroek}, Han and {Deng}, Yuanyong and {Li}, Xin and {Yang}, Xiao and {Zhang}, Mei},
        title = "{Observations of pores and surrounding regions with CO 4.66 {\ensuremath{\mu}}m lines by BBSO/CYRA}",
      journal = {\aap},
         year = 2023,
        month = jan,
       volume = {669},
          eid = {A79},
        pages = {A79},
          doi = {10.1051/0004-6361/202244600},
archivePrefix = {arXiv},
       eprint = {2211.07100},
 primaryClass = {astro-ph.SR},
       adsurl = {https://ui.adsabs.harvard.edu/abs/2023A&A...669A..79S}
}

@ARTICLE{Stauffer2022,
       author = {{Stauffer}, Johnathan R. and {Reardon}, Kevin P. and {Penn}, Matt},
        title = "{Chromospheric Carbon Monoxide Formation around a Solar Pore}",
      journal = {\apj},
         year = 2022,
        month = may,
       volume = {930},
       number = {1},
          eid = {87},
        pages = {87},
          doi = {10.3847/1538-4357/ac59b0},
       adsurl = {https://ui.adsabs.harvard.edu/abs/2022ApJ...930...87S}
}

@software{tfbi_theory_and_SymSolver,
  author       = {Evans, Samuel},
  title        = {tfbi\_theory + SymSolver},
  month        = jan,
  year         = 2026,
  publisher    = {Zenodo},
  version      = {2025.6.0},
  doi          = {10.5281/zenodo.18381591},
  url          = {https://doi.org/10.5281/zenodo.18381591},
}

@ARTICLE{Wargnier2022,
       author = {{Wargnier}, Q.~M. and {Mart{\'\i}nez-Sykora}, J. and {Hansteen}, V.~H. and {De Pontieu}, B.},
        title = "{Detailed Description of the Collision Frequency in the Solar Atmosphere}",
      journal = {\apj},
         year = 2022,
        month = jul,
       volume = {933},
       number = {2},
          eid = {205},
        pages = {205},
          doi = {10.3847/1538-4357/ac6e62},
       adsurl = {https://ui.adsabs.harvard.edu/abs/2022ApJ...933..205W}
}

@ARTICLE{Wedemeyer2004,
       author = {{Wedemeyer}, S. and {Freytag}, B. and {Steffen}, M. and {Ludwig}, H. -G. and {Holweger}, H.},
        title = "{Numerical simulation of the three-dimensional structure and dynamics of the non-magnetic solar chromosphere}",
      journal = {\aap},
         year = 2004,
        month = feb,
       volume = {414},
        pages = {1121-1137},
          doi = {10.1051/0004-6361:20031682},
archivePrefix = {arXiv},
       eprint = {astro-ph/0311273},
 primaryClass = {astro-ph},
       adsurl = {https://ui.adsabs.harvard.edu/abs/2004A&A...414.1121W}
}

@ARTICLE{Wiltberger2017,
       author = {{Wiltberger}, M. and {Merkin}, V. and {Zhang}, B. and {Toffoletto}, F. and {Oppenheim}, M. and {Wang}, W. and {Lyon}, J.~G. and {Liu}, J. and {Dimant}, Y. and {Sitnov}, M.~I. and {Stephens}, G.~K.},
        title = "{Effects of electrojet turbulence on a magnetosphere-ionosphere simulation of a geomagnetic storm}",
      journal = {J. Geophys. Res.},
         year = 2017,
        month = may,
       volume = {122},
       number = {5},
        pages = {5008-5027},
          doi = {10.1002/2016JA023700},
       adsurl = {https://ui.adsabs.harvard.edu/abs/2017JGRA..122.5008W}
}

@ARTICLE{ZhangVarney2024,
       author = {Zhang, Yizhe and Varney, Roger H.},
        title = {A Statistical Survey of E-Region Anomalous Electron Heating Using Poker Flat Incoherent Scatter Radar Observations},
      journal = {Journal of Geophysical Research: Space Physics},
       volume = {129},
       number = {9},
        pages = {e2023JA032360},
          doi = {https://doi.org/10.1029/2023JA032360},
          url = {https://agupubs.onlinelibrary.wiley.com/doi/abs/10.1029/2023JA032360},
       eprint = {https://agupubs.onlinelibrary.wiley.com/doi/pdf/10.1029/2023JA032360},
         note = {e2023JA032360 2023JA032360},
         year = {2024}
}

@ARTICLE{Zaqarashvili2011,
       author = {{Zaqarashvili}, T.~V. and {Khodachenko}, M.~L. and {Rucker}, H.~O.},
        title = "{Damping of Alfv{\'e}n waves in solar partially ionized plasmas: effect of neutral helium in multi-fluid approach}",
      journal = {\aap},
         year = 2011,
        month = oct,
       volume = {534},
          eid = {A93},
        pages = {A93},
          doi = {10.1051/0004-6361/201117380},
archivePrefix = {arXiv},
       eprint = {1109.1154},
 primaryClass = {astro-ph.SR},
       adsurl = {https://ui.adsabs.harvard.edu/abs/2011A&A...534A..93Z}
}

@ARTICLE{Khomenko2014,
       author = {{Khomenko}, E. and {Collados}, M. and {D{\'\i}az}, A. and {Vitas}, N.},
        title = "{Fluid description of multi-component solar partially ionized plasma}",
      journal = {Physics of Plasmas},
         year = 2014,
        month = sep,
       volume = {21},
       number = {9},
          eid = {092901},
        pages = {092901},
          doi = {10.1063/1.4894106},
archivePrefix = {arXiv},
       eprint = {1408.1871},
 primaryClass = {astro-ph.SR},
       adsurl = {https://ui.adsabs.harvard.edu/abs/2014PhPl...21i2901K}
}

@ARTICLE{Shelyag2016,
       author = {{Shelyag}, S. and {Khomenko}, E. and {de Vicente}, A. and {Przybylski}, D.},
        title = "{Heating of the Partially Ionized Solar Chromosphere by Waves in Magnetic Structures}",
      journal = {\apjl},
         year = 2016,
        month = mar,
       volume = {819},
       number = {1},
          eid = {L11},
        pages = {L11},
          doi = {10.3847/2041-8205/819/1/L11},
archivePrefix = {arXiv},
       eprint = {1602.03373},
 primaryClass = {astro-ph.SR},
       adsurl = {https://ui.adsabs.harvard.edu/abs/2016ApJ...819L..11S}
}

@article{young_hybrid_2017,
	title = {Hybrid simulations of coupled {Farley}-{Buneman}/gradient drift instabilities in the equatorial E region ionosphere},
	volume = {122},
	doi = {10.1002/2017ja024161},
	number = {5},
	journal = {Journal of Geophysical Research: Space Physics},
	author = {Young, Matthew A. and Oppenheim, Meers M. and Dimant, Yakov S.},
	month = may,
	year = {2017},
	note = {Publisher: American Geophysical Union (AGU)},
	pages = {5768--5781},
}

@book{butcher_numerical_2016,
	address = {Newark, UNITED KINGDOM},
	title = {Numerical {Methods} for {Ordinary} {Differential} {Equations}},
	isbn = {978-1-119-12151-0},
	url = {http://ebookcentral.proquest.com/lib/bu/detail.action?docID=4591869},
	publisher = {John Wiley \& Sons, Incorporated},
	author = {Butcher, J. C.},
	year = {2016},
}

@article{Brooks2015,
    adsurl = {https://ui.adsabs.harvard.edu/abs/2015NatCo...6.5947B},
    archiveprefix = {arXiv},
    author = {{Brooks}, David H. and {Ugarte-Urra}, Ignacio and {Warren}, Harry P.},
    doi = {10.1038/ncomms6947},
    eid = {5947},
    eprint = {1605.09514},
    journal = {Nature Communications},
    month = jan,
    pages = {5947},
    primaryclass = {astro-ph.SR},
    title = {{Full-Sun observations for identifying the source of the slow solar wind}},
    volume = {6},
    year = 2015
}

@article{Laming2015,
    adsurl = {http://adsabs.harvard.edu/abs/2015LRSP...12....2L},
    archiveprefix = {arXiv},
    author = {{Laming}, J.~M.},
    doi = {10.1007/lrsp-2015-2},
    eid = {2},
    eprint = {1504.08325},
    journal = {Living Reviews in Solar Physics},
    month = sep,
    pages = {2},
    primaryclass = {astro-ph.SR},
    title = {{The FIP and Inverse FIP Effects in Solar and Stellar Coronae}},
    volume = 12,
    year = 2015
}

@article{Laming2017,
    adsurl = {http://adsabs.harvard.edu/abs/2017ApJ...844..153L},
    archiveprefix = {arXiv},
    author = {{Laming}, J.~M.},
    doi = {10.3847/1538-4357/aa7cf1},
    eid = {153},
    eprint = {1707.05378},
    journal = {\apj},
    month = aug,
    pages = {153},
    primaryclass = {astro-ph.SR},
    title = {{The First Ionization Potential Effect from the Ponderomotive Force: On the Polarization and Coronal Origin of Alfv{\'e}n Waves}},
    volume = 844,
    year = 2017
}

@article{Long2024,
    adsurl = {https://ui.adsabs.harvard.edu/abs/2024ApJ...965...63L},
    archiveprefix = {arXiv},
    author = {{Long}, David M. and {Baker}, Deborah and {To}, Andy S.~H. and {van Driel-Gesztelyi}, Lidia and {Brooks}, David H. and {Stangalini}, Marco and {Murabito}, Mariarita and {James}, Alexander W. and {Mathioudakis}, Mihalis and {Testa}, Paola},
    doi = {10.3847/1538-4357/ad3234},
    eid = {63},
    eprint = {2403.06711},
    journal = {\apj},
    month = apr,
    number = {1},
    pages = {63},
    primaryclass = {astro-ph.SR},
    title = {{Identifying Plasma Fractionation Processes in the Chromosphere Using IRIS}},
    volume = {965},
    year = 2024}

@article{MartinezSykora2023,
    adsurl = {https://ui.adsabs.harvard.edu/abs/2023ApJ...949..112M},
    archiveprefix = {arXiv},
    author = {{Mart{\'\i}nez-Sykora}, Juan and {De Pontieu}, Bart and {Hansteen}, Viggo H. and {Testa}, Paola and {Wargnier}, Q.~M. and {Szydlarski}, Mikolaj},
    doi = {10.3847/1538-4357/acc465},
    eid = {112},
    eprint = {2211.09361},
    journal = {\apj},
    month = jun,
    number = {2},
    pages = {112},
    primaryclass = {astro-ph.SR},
    title = {{The Impact of Multifluid Effects in the Solar Chromosphere on the Ponderomotive Force under SE and NEQ Ionization Conditions}},
    volume = {949},
    year = 2023
}

@article{Wargnier2023,
    adsurl = {https://ui.adsabs.harvard.edu/abs/2023ApJ...946..115W},
    archiveprefix = {arXiv},
    author = {{Wargnier}, Q.~M. and {Mart{\'\i}nez-Sykora}, J. and {Hansteen}, V.~H. and {De Pontieu}, B.},
    doi = {10.3847/1538-4357/acbfb1},
    eid = {115},
    eprint = {2211.02157},
    journal = {\apj},
    month = apr,
    number = {2},
    pages = {115},
    primaryclass = {astro-ph.SR},
    title = {{Multifluid Simulations of Upper-chromospheric Magnetic Reconnection with Helium-Hydrogen Mixture}},
    volume = {946},
    year = 2023
}

@ARTICLE{Testa2010,
       author = {{Testa}, Paola},
        title = "{X-ray emission processes in stars and their immediate environment}",
      journal = {Proceedings of the National Academy of Science},
         year = 2010,
        month = apr,
       volume = {107},
       number = {16},
        pages = {7158-7163},
          doi = {10.1073/pnas.0913822107},
archivePrefix = {arXiv},
       eprint = {1008.4343},
 primaryClass = {astro-ph.SR},
       adsurl = {https://ui.adsabs.harvard.edu/abs/2010PNAS..107.7158T}
}

@article{Testa2015,
    adsurl = {http://adsabs.harvard.edu/abs/2015RSPTA.37340259T},
    archiveprefix = {arXiv},
    author = {{Testa}, P. and {Saar}, S.~H. and {Drake}, J.~J.},
    doi = {10.1098/rsta.2014.0259},
    eprint = {1502.07401},
    journal = {Philosophical Transactions of the Royal Society of London Series A},
    month = apr,
    pages = {20140259-20140259},
    primaryclass = {astro-ph.SR},
    title = {{Stellar activity and coronal heating: an overview of recent results}},
    volume = 373,
    year = 2015
}

@article{Testa2023,
    adsurl = {https://ui.adsabs.harvard.edu/abs/2023ApJ...944..117T},
    archiveprefix = {arXiv},
    author = {{Testa}, Paola and {Mart{\'\i}nez-Sykora}, Juan and {De Pontieu}, Bart},
    doi = {10.3847/1538-4357/acb343},
    eid = {117},
    eprint = {2211.07755},
    journal = {\apj},
    month = feb,
    number = {2},
    pages = {117},
    primaryclass = {astro-ph.SR},
    title = {{Coronal Abundances in an Active Region: Evolution and Underlying Chromospheric and Transition Region Properties}},
    volume = {944},
    year = 2023
}

%%% ---------------- APPENDIX ---------------- %%%
\appendix

\section{Simulation Highlight --- ``Standing Wave'' After Turbulence}
  \label{sec:appendix:highlight_standing_wave}

  This section highlights one simulation with unusual behavior: a standing wave pattern develops \emph{after} the turbulent regime, in simulation (3,2,3,2,4)F.

  \begin{figure}[htbp]
    \centering
    \includegraphics[width=0.85\textwidth]{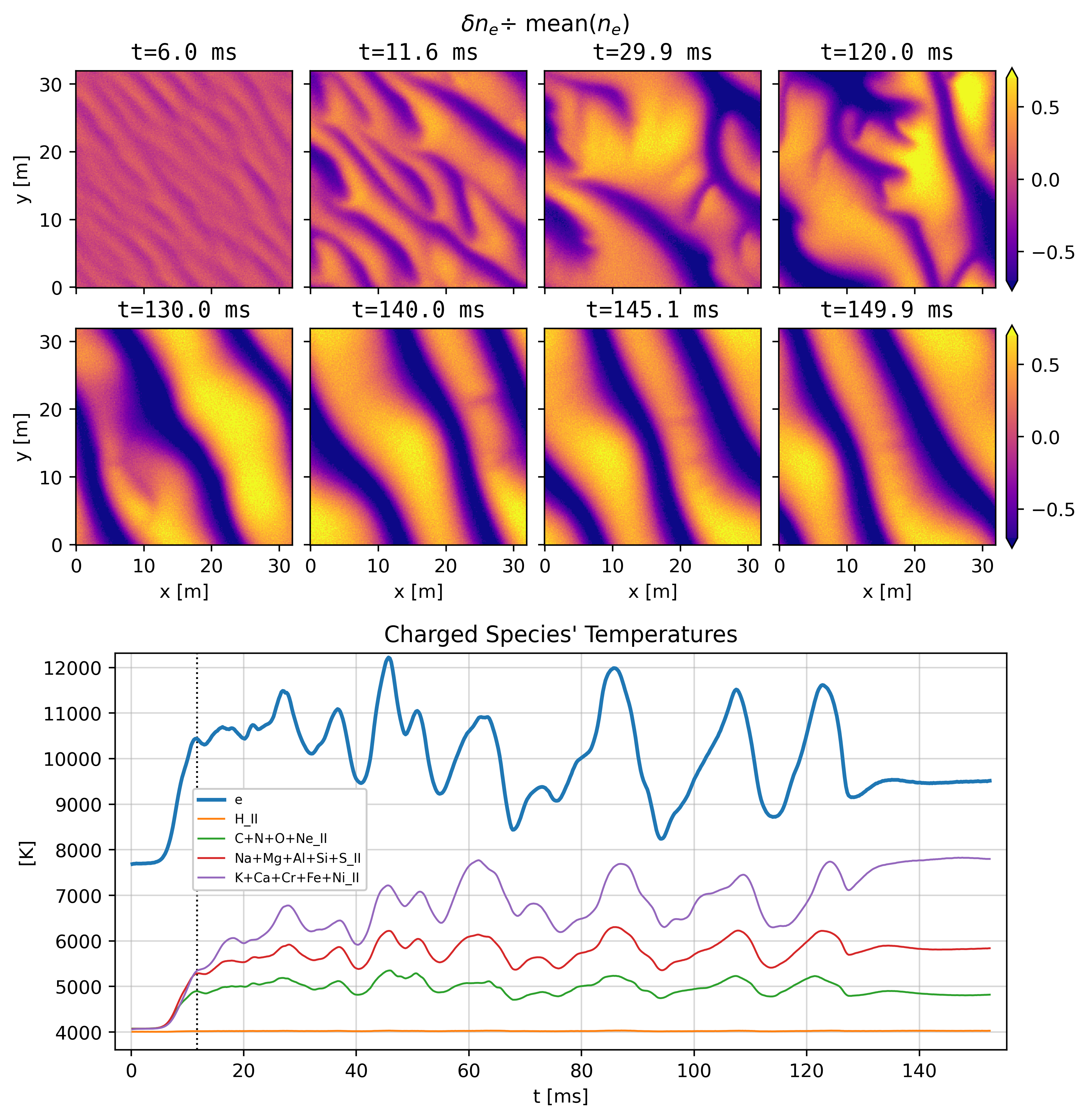}
    \caption{
      Electron density perturbations relative to the mean, and all charged species' average temperatures throughout simulation (3,2,3,2,4)F.
      Temperature plot has the same formatting as panel~(C) of Figure~\ref{fig:typical_sim_megaplot}.
      The vertical dashed line marks $t\textsub{turb}$ for this simulation.
    }
    \label{fig:highlight_standing_wave}
  \end{figure}

  Figure~\ref{fig:highlight_standing_wave} shows electron density perturbations and temperatures throughout simulation (3,2,3,2,4)F.
  The post-turbulence standing wave pattern starts to develop around $t{=}130\uu{ms}$.
  Before that, panels at the top of the figure show
    a well-resolved linear regime around $t{=}6.0\uu{ms}$, with more than 12 waves across the box;
    turbulent features around $t{=}11.6\uu{ms}$ (the closest saved snapshot time to $t\textsub{turb}{=}11.72\uu{ms}$);
    larger turbulent structures around $t{=}29.9\uu{ms}$, with only a few large-scale structures across the box;
    and that a ``quasi-steady'' turbulent regime lasts until around $t{=}120\uu{ms}$, with qualitatively similar features as in the previous panel.
  During the quasi-steady turbulent regime, structures change over time, though the number of large and small scale features in the box remains roughly the same.
  Additionally, charged species' temperatures fluctuate around some average value, for example $T_{e}$ fluctuates around roughly $10500\uu{K}$.
  
  After the standing wave pattern develops, density structures mostly stop changing, instead just flowing across the box, and temperature fluctuations almost entirely cease.
  The direction of flow is predominantly leftwards, with some amount of downward motion.
  For example, the density peak near $(x,y)=$~(15m,~15m) at $t{=}130\uu{ms}$ moves to roughly $(x,y)=$~(4m,~11m) by $t{=}149.9\uu{ms}$.
  When the temperature fluctuations settle, the electron temperature is roughly $9500\uu{K}$, which is $1000\uu{K}$ less than the average value it was fluctuating around before the standing wave pattern developed.
  Meanwhile, the post-fluctuations temperature of \iII{K+Ca+Cr+Fe+Ni} ends up roughly $800\uu{K}$ more than the average value it was fluctuating around.

  The cause of this standing wave pattern remains unclear.
  The finite box size coupled with periodic boundary conditions might play an important role.
  However, when we performed a simulation using the same physical conditions but doubled the box length in each direction, we saw a similar standing wave pattern develop, also at around $t{=}130\uu{ms}$, and with features of similar shapes and physical size.
  If this pattern is indeed physically realizable, and not just a numerical effect, it might be an interesting topic for future study.

\section{Simulation Highlight ---  TFBI-driven Cooling}
  \label{sec:appendix:highlight_cooling}
  
  This section highlights one simulation with unusual behavior: average electron temperatures \emph{decrease} due to turbulence, in simulation (4,3,5,2,4)G.

  \begin{figure}[htbp]
    \centering
    \includegraphics[width=0.85\textwidth]{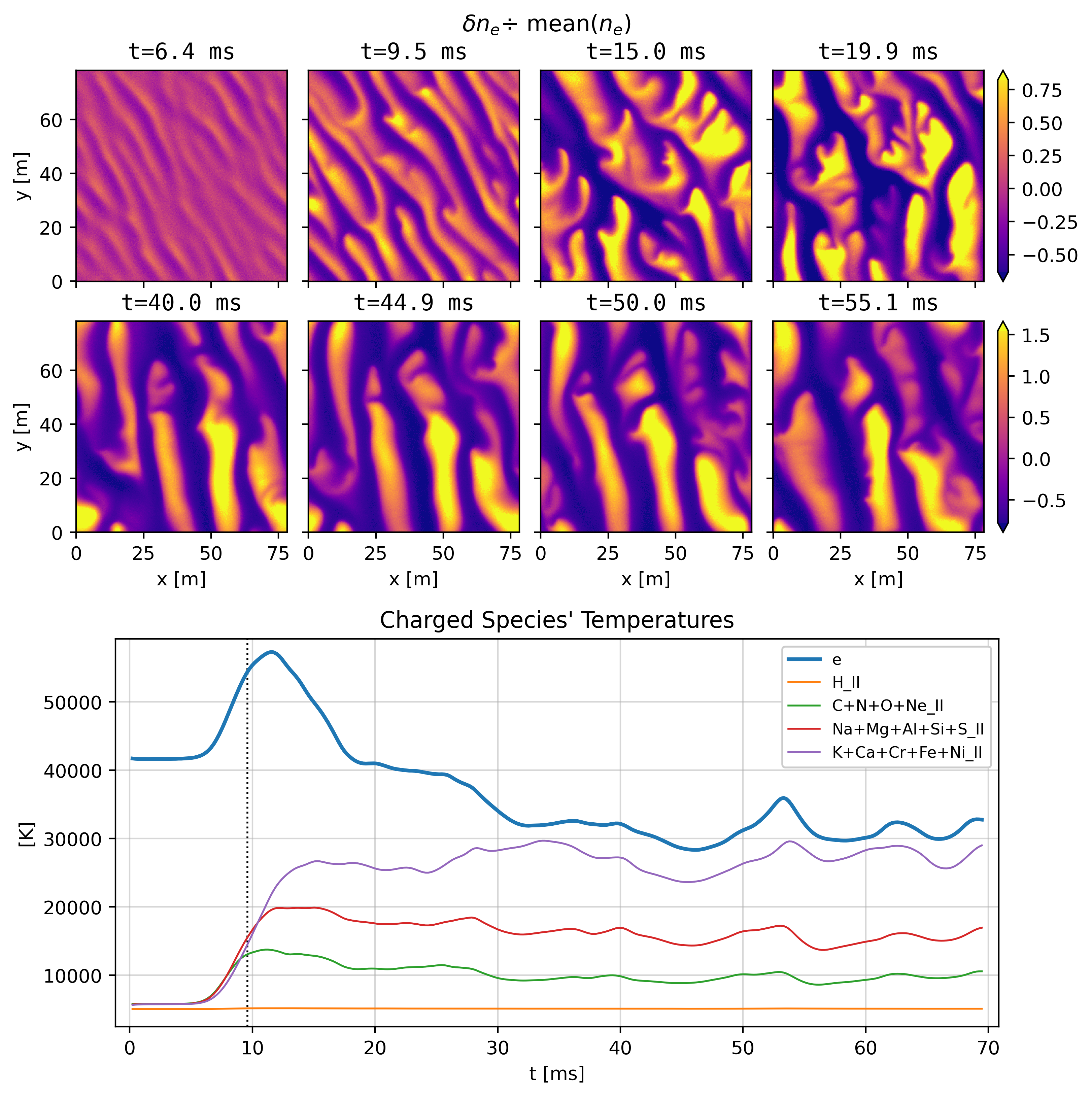}
    \caption{
      Electron density perturbations relative to the mean, and all charged species' average temperatures throughout simulation (4,3,5,2,4)G, with the same plot formatting as in Figure~\ref{fig:highlight_standing_wave}.
    }
    \label{fig:highlight_turbulent_cooling}
  \end{figure}

  Figure~\ref{fig:highlight_turbulent_cooling} shows electron density perturbations and temperatures throughout simulation (4,3,5,2,4)G.
  The simulation initially exhibits ``usual'' behavior in the turbulent regime, with turbulent structures evolving and charged species' temperatures increasing.
  However, the electron temperatures then cool significantly before fluctuating around a value roughly $10000\uu{K}$ colder than $T_{e}\oZ$.
  The electron density perturbation doesn't show anything particularly unusual at first glance, having
    a well-resolved linear regime around $t{=}6.4\uu{ms}$, with roughly 14 waves across the box;
    turbulent features around $t{=}9.5\uu{ms}$ (the closest saved snapshot time to $t\textsub{turb}{=}9.60\uu{ms}$);
    and scale size of turbulent structures increasing over time, until roughly $t{=}40\uu{ms}$.
  A closer inspection of times after $t{\approx}40\uu{ms}$ reveals that the structures evolve less rapidly and their flow across the box slows down.
  For example, the density enhancement around $x{=}55\uu{m}$, $y{<}40\uu{m}$, $t{=}40\uu{ms}$ retains a similar size and shape throughout the next 15\uu{ms}, while flowing to roughly $x{=}40\uu{m}$ by $t{=}55.1\uu{ms}$.
  In the earlier snapshots, such as between $t{=}15.0\uu{ms}$ and $t{=}19.9\uu{ms}$, features tend to evolve significantly over timescales of only a few ms.
  
  This simulation, and the one other simulation in the TFBI suite with the largest $E/B$~speed ($E\oZ/B=10^{4.5}\uu{m/s}$), which also predicts turbulent cooling of electrons, might be physically inaccurate due to neglecting radiative effects, which might become significant at these electron temperatures.
  Additionally, electron temperatures in some grid cells of the simulation box itself become hotter than $10^{5}\uu{K}$, but only in low density regions.
  If these hot electrons appeared in the actual solar chromosphere, their interactions with neutral atoms might enable many energy transitions and heavily influence observations.
  However, their effect on local energy transfer mediated by collisions (i.e., excluding radiation physics) seems to be small; the average inferred neutral heating rate still decreased by roughly 20\% (as indicated by the point marked with an ``x'' in Figure~\ref{fig:dTndt_vs_Erel}, around ${(E\oZ/E\textsub{thr})-1}=3.8$), when compared to its zeroth-order value.
  The plasma is mostly neutral, so the neutral heating rates give good insight into the amount of energy being converted from flows into heating.

\section{Simulation Highlight --- Ion-Ion Soliton Mediated by Turbulence?}
  \label{sec:appendix:highlight_soliton}

  This section highlights one simulation with unusual behavior: simulation (3,1,4,4,6)E shows a soliton-like feature in ion densities, but not electron densities, which persists over long periods of time while propagating slowly, on top of the turbulent background.

  \begin{figure}[htbp]
    \centering
    \includegraphics[width=\textwidth]{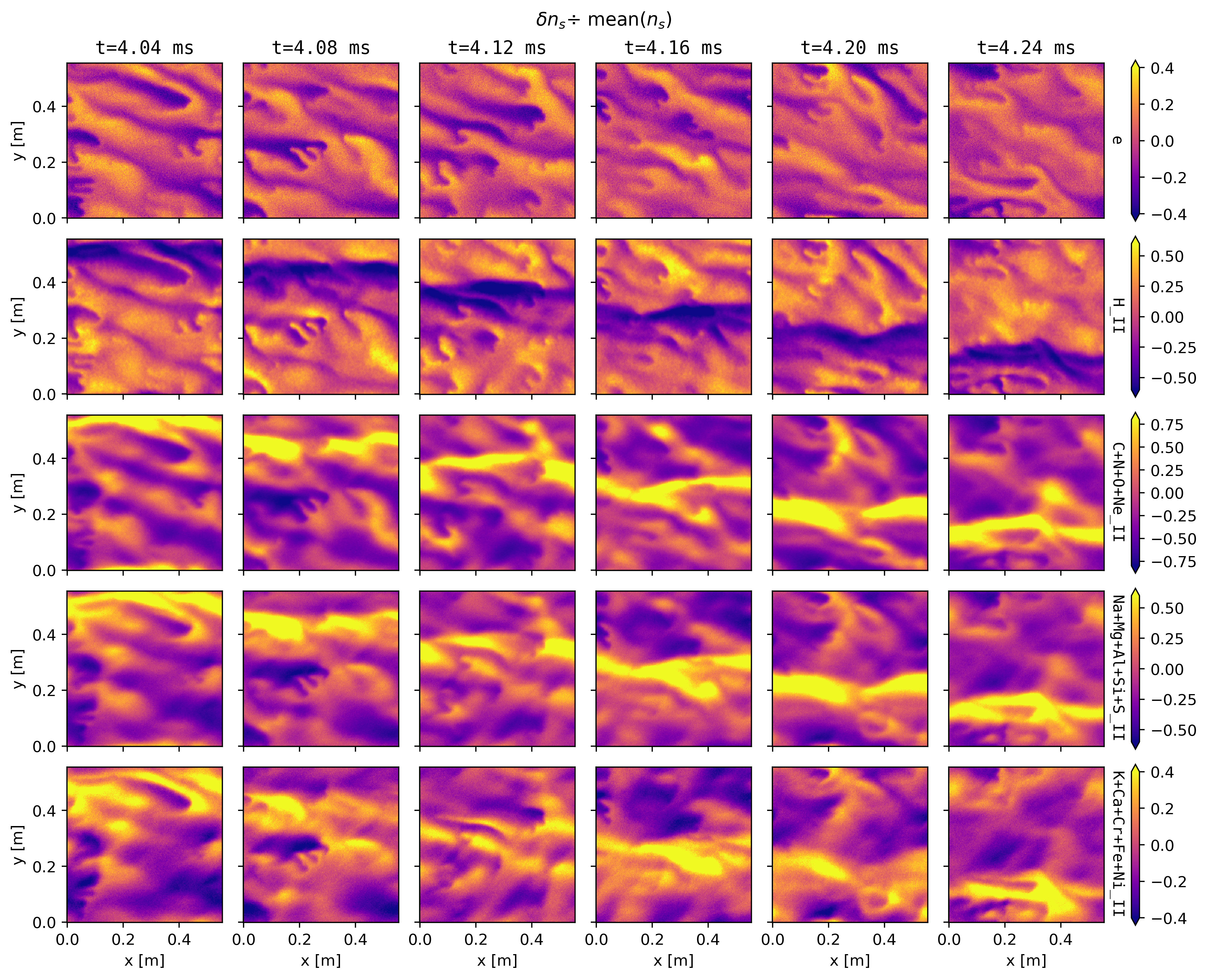}
    \caption{
      Density perturbations relative to the mean, at selected snapshots of simulation (3,1,4,4,6)E.
      From subplot to subplot, time increases from left to right (as indicated by labels on top), while each row shows a different charged species (as indicated by labels on the right).
    }
    \label{fig:highlight_soliton}
  \end{figure}

  Figure~\ref{fig:highlight_soliton} shows density perturbations of all charged species throughout simulation (3,1,4,4,6)E.
  There is a density depletion in \iII{H} located cospatially with density enhancements in all heavy ions, and no noticeable enhancement nor depletion in electron density.
  This feature propagates downwards at roughly 0.4\uu{m} per 20\uu{ms}, much slower than the motions of smaller-scale turbulent features across the box.
  Small-scale turbulent features continually crash against this large stable feature without significantly disturbing it.
  Much earlier in the simulation (not shown here), by around $t{=}1\uu{ms}$, two similar, large-scale, but somewhat weaker features appear and propagate with roughly the same velocity, but are disrupted somewhat around $t{=}2.3\uu{ms}$, then merge into a single feature at roughly $t{=}2.8\uu{ms}$.
  That single feature persists until the end of the simulation at $t{=}4.86\uu{ms}$, and the figure shows its motion from near the top of the simulation box at $t{=}4.04\uu{ms}$, to near the bottom of the simulation box at $t{=}4.24\uu{ms}$.

  Note that the \iII{C+N+O+Ne} largely acts as a tracer in this simulation, where $n_{i}\oZ/n_{e}\oZ = 0.500$, $5.18\times10^{-7}$, $0.334$, and $0.166$ for $i{=}$\iII{H}, \iII{C+N+O+Ne}, \iII{Na+Mg+Al+Si+S}, and \iII{K+Ca+Cr+Fe+Ni}, respectively.
  It is unclear whether this feature results from numerical effects such as the periodic and finite simulation box, or represents the discovery of a new, self-reinforcing wave structure (i.e., a soliton), mediated by TFBI-driven turbulence.
  We leave a more detailed investigation of this feature to future work.

\section{TFBI in Hybrid EPPIC When Missing From Pure-PIC}
  \label{sec:appendix:hybrid_tfbi}

  This section discusses the points where the purely kinetic version of EPPIC failed to generate TFBI growth, as well as one such point where a hybrid EPPIC simulation does in fact show instability growth and turbulence.

  We simulated 10 additional points, excluded from the TFBI suite (Table~\ref{tab:tfbi_suite_params}), where the TFBI theory solver predicts growth but purely kinetic EPPIC fails to generate the TFBI.
  The 5-indexes of these points are: (2,0,2,3,4), (2,0,2,3,6), (2,0,3,3,6), (2,0,3,4,6), (2,1,4,3,4), (2,1,4,3,5), (2,1,4,3,6), (2,1,4,4,6), (2,2,4,3,6), and (2,3,4,3,6).

  One possible reason kinetic EPPIC might fail to develop TFBI growth is excessive background noise levels due to the finite number of PIC particles.
  In simulations that successfully show growth, density perturbation strength grows larger than the background noise level then eventually saturates, with saturation level varying across simulations.
  If the saturation level were instead comparable to or smaller than the PIC particle noise level, the noise could prevent TFBI growth.
  Another possibility is the TFBI theory solver excluding kinetic effects which cause damping in EPPIC, such that the solver significantly underestimates wavelengths or overestimates growth rates, causing the corresponding simulations to be much too small or end much too early to produce the TFBI.
  We attempted three separate runs at one such point (with index (2,1,4,3,6)), using 16 times more PIC particles, doubling the spatial extent of the simulation, and running the simulation for five times longer, but none of these led to TFBI growth.

  However, when simulating point (2,1,4,3,6) with a newly-developed hybrid version of EPPIC, we managed to see an instability grow.
  Hybrid EPPIC is an extension of EPPIC that models ion dynamics with the traditional PIC method as described in Section~\ref{sec:methods:eppic_simulator}, but makes the approximation that electrons are represented by an intertialess fluid.
  We use a new version of hybrid EPPIC similar to the original version \citep{young_hybrid_2017} but here including a thermal equation for electrons rather than using an isothermal or adiabatic equation of state.

  The electron thermal equation in hybrid EPPIC is given by:
  \begin{equation}\label{eq:thermal_electron}
  	% \pdtime{T_{e}} = -\vec{u}_{e}\cdot\grad{T_{e}} - \frac{2}{3}T_e\nabla\cdot\vec{u}_{e} + \frac{2}{3}\mu_{e,n}\nu_{e,n}|\vec{u}_{e}|^2 - \delta_{e,n}\nu_{e,n}(T_{e} - T_n)  % extra spacing
    \pdtime{T_{e}}=-\vec{u}_{e}\cdot\grad{T_{e}}-\frac{2}{3}T_e\nabla\cdot\vec{u}_{e}+\frac{2}{3}\mu_{e,n}\nu_{e,n}|\vec{u}_{e}|^2-\delta_{e,n}\nu_{e,n}(T_{e}-T_n)  % apj spacing
  \end{equation}

  where $\mu_{e,n}$ is the electron reduced mass, $\nu_{e,n}$ is the electron neutral collision frequency, and $\delta_{e,n}$ represents the fraction of energy lost by electrons when colliding with neutrals, as described in \citep{Dimant2004}.
  Note that this equation is equivalent to equation~\eqref{eq:heating} for electrons in the case of elastic collisions, as relevant to the chromosphere.
  To solve equation \eqref{eq:thermal_electron} we first find the electron velocity using equation~\eqref{eq:momentum} with the left-side term set to 0.
  We then apply a second-order Adams-Bashforth predictor-corrector method \citep{butcher_numerical_2016} to update the electron temperature using equation \eqref{eq:thermal_electron}.
  The electron temperature is then used in the usual field solve as described in \citep{young_hybrid_2017}.

  \begin{figure}[htbp]
  	\centering
  	\includegraphics[width=0.85\textwidth]{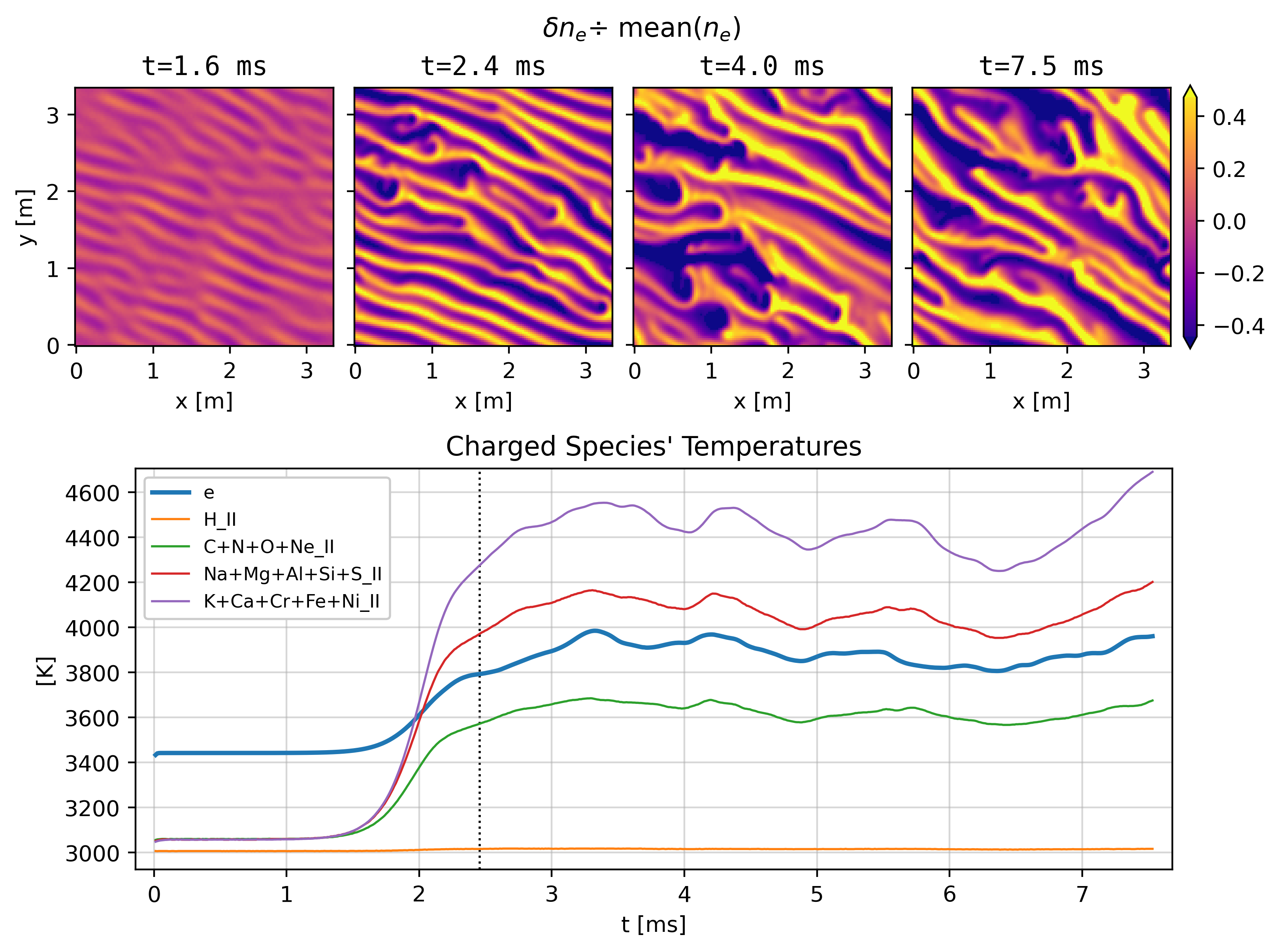}
  	\caption{
  		Electron density perturbations relative to the mean (top), and all charged species' average temperatures (bottom) throughout the (2,1,4,3,6) hybrid simulation.
  		Temperature plot has the same formatting as Panel~(C) of Figure~\ref{fig:typical_sim_megaplot}.
  		The vertical dashed line marks $t\textsub{turb}$ for this simulation.
  	}
  	\label{fig:hybrid_eppic_tfbi}
  \end{figure}

  Figure~\ref{fig:hybrid_eppic_tfbi} shows electron densities and temperatures in the hybrid simulation of (2,1,4,3,6), which successfully produces TFBI growth and turbulence.
    The densities show a well-resolved linear regime around $t{=}1.6\uu{ms}$ with roughly 15 waves across the box, followed by the development of some turbulent features by $t{=}2.45\uu{ms}$ (the closest saved snapshot to $t\textsub{turb}{=}2.46\uu{ms}$).
    More turbulent features develop, and the scale size of features increases until the end of the simulation at $t{=}7.53\uu{ms}$.
    The temperatures exhibit typical behavior, starting near their equilibrium $T_{s}\oZ$ values, increasing due to turbulence, then fluctuating.
    Applying the methods described in Section~\ref{sec:results:typical_tfbi}, we compute turbulent heatings of 
      $\Delta{T}_{e}\textsup{turb}=443\pm50.\uu{K}$, 
      $\Delta{T}\subII{H}\textsup{turb}=9.0\pm1.2\uu{K}$,   
      $\Delta{T}\subII{C+N+O+Ne}\textsup{turb}=567\pm33\uu{K}$,  
      $\Delta{T}\subII{Na+Mg+Al+Si+S}\textsup{turb}=1011\pm61\uu{K}$ and
      $\Delta{T}\subII{K+Ca+Cr+Fe+Ni}\textsup{turb}=1378\pm94\uu{K}$.

  We compare these turbulent heatings to values predicted using the empirical relationships from Figure~\ref{fig:heating_vs_Erel}.
    Plugging in the relative surplus driving field at this point, $(E\oZ/E\textsub{thr})-1=0.51$, the predicted amounts of turbulent heating would be:
      $\Delta{T}_{e}\textsup{pred}=586\pm155\uu{K}$,
      $\Delta{T}\subII{H}\textsup{pred}=-11.5\pm6.2\uu{K}$,
      $\Delta{T}\subII{C+N+O+Ne}\textsup{pred}=-171\pm167\uu{K}$,
      $\Delta{T}\subII{Na+Mg+Al+Si+S}\textsup{pred}=39\pm192\uu{K}$ and
      $\Delta{T}\subII{K+Ca+Cr+Fe+Ni}\textsup{pred}=829\pm286\uu{K}$.
    For electrons and \iII{K+Ca+Cr+Fe+Ni}, predictions agree with simulated heating to within one and two standard deviations, respectively.
    For all other species, the simulated values are more than three standard deviations larger than predicted.
  
  There are a few possible explanations for these discrepancies.
    As mentioned in Section~\ref{sec:discussion}, trends with relative surplus driving field may become nonlinear near the threshold field.
    Thus, one should not necessarily expect trends to hold when extrapolating down to $E\oZ/E\textsub{thr}=1.51$ as it occurs beyond the smallest value in the TFBI suite of $E\oZ/E\textsub{thr}=1.70$.
    Additionally, hybrid EPPIC neglects kinetic electron effects, which may play a role in regulating ion temperatures.
    However, it also does not have any artifical density scaling effects, which likely suppress the instability across the TFBI simulation suite.
  
  We emphasize that the simulated heating in this hybrid EPPIC simulation is still consistent with the main results of this work as long as those results are treated as providing \emph{lower bounds} on TFBI-driven heating.
    Although hybrid EPPIC excludes kinetic electron effects, its ability to use unscaled densities and to simulate the TFBI at smaller values of relative surplus driving field provides a promising avenue for future studies of TFBI turbulence.

\section{Reproducibility}
  \label{sec:appendix:repro}
  To facilitate reproduction of these results, we provide the relevant simulation data and Python analysis routines online with DOI:
    (TBD. Note: we would create this immediately after the review process, but before publication.) %, to avoid managing multiple versions corresponding to changes made during review.)
  This includes
    all relevant input decks and simulation snapshots from all TFBI suite simulations and the hybrid EPPIC simulation from Appendix~\ref{sec:appendix:hybrid_tfbi}.

\end{document}